%% file: main.tex
\documentclass[runningheads,orivec,sts]{llncs}

\usepackage{microtype}
\usepackage{xspace}
\usepackage{times}
\usepackage{subfig}
\usepackage{textgreek}
\usepackage{latexsym,amssymb,amsmath,amsfonts}
\usepackage[defblank]{paralist}
\usepackage{tikz}
\usetikzlibrary{arrows,shapes,snakes,automata,backgrounds,petri,positioning,shadows,matrix,decorations.pathmorphing,fit,positioning,calc,backgrounds}
\usepackage{todonotes}
\usepackage{url}
\usepackage{hyperref}
\usepackage{xcolor}

\usepackage[scaled=0.92]{newtxtt} % nicer mono font
\usepackage{listings,lstautogobble}
\usepackage{graphicx}

\usepackage{multirow}
\usepackage{booktabs} 

\usepackage{comment}

\input{macros}

\title{Modeling and In-Database Management of \\Relational, Data-Aware Processes (Extended Version)}

\author{Diego Calvanese\inst{1} \and Marco Montali\inst{1} \and Fabio Patrizi\inst{2} \and
 Andrey Rivkin\inst{1}}
\authorrunning{Calvanese, Montali, Patrizi, Rivkin} \institute{
 Free University of Bozen-Bolzano,
 Piazza Domenicani 3, 39100 Bolzano, Italy\\
 \email{calvanese,montali,rivkin@inf.unibz.it}
 \and Sapienza Universit\`a di Roma\\
  \email{patrizi@dis.uniroma1.it}
}

\begin{document}

\maketitle

\begin{abstract}
  \input{abstract}
\end{abstract}

\input{introduction}
\input{daphne-model}

\input{daphne-framework}
\input{discussion-and-related-work}

\input{conclusion}

\bibliographystyle{splncs04}
\bibliography{main}
%\bibliography{string-tiny,local}

%%%%%%%%%%%%%%%%%%%%%%%%%%%%%%
%%%%%%%%%%		APPENDIX	 		%%%%%%%%%%
%%%%%%%%%%%%%%%%%%%%%%%%%%%%%%
\clearpage
\appendix

\pagenumbering{roman}

\input{appendix}

\end{document}

%% file: macros.tex
\makeatletter
\newcommand{\nobrackettag}[0]{\def\tagform@##1{\maketag@@@{##1}}}
\makeatother
\newcommand{\qedtriangle}{\hfill$\triangleleft$}

\usepackage{xspace}
\usepackage{array}
\usepackage{bbold}
\usepackage{mathastext}

\newcommand{\proj}[2]{\ensuremath{\relname{#1}[#2]}}

\newcommand{\FK}[4]{\proj{#1}{#2}\rightarrow\proj{#3}{#4}}
\newcommand{\pk}[1]{\textsc{pk}(\relname{#1})}
\newcommand{\rid}{\textsf{RID}}
\newcommand{\hash}{\textsf{hash}}
\newcommand{\logstate}{\textsf{state}}
\newcommand{\id}{\textsf{ID}}

\newcommand{\set}[1]{\{#1\}}
\newcommand{\tup}[1]{\langle#1\rangle}

\newcommand{\daphne}{{\textsc{daphne}}\xspace}

\newcommand{\dapm}{\textit{DAPS}\xspace}
\newcommand{\dapsl}{\textbb{dapSL}\xspace}

\newcommand{\dbengine}{{\small{\textbf{DB Engine}}}\xspace}
\newcommand{\fengine}{{\small{\textbf{Flow Engine}}}\xspace}
\newcommand{\smanager}{{\small{\textbf{Service Manager}}}\xspace}

\xspaceaddexceptions{\_}
\xspaceaddexceptions{\,}
\newcommand{\action}{\textalpha\xspace}
\newcommand{\state}{\textsf{s}\xspace}
\newcommand{\binding}{\textsf{b}\xspace}
\newcommand{\caeval}{\texttt{\action\_ca\_eval(\state)}\xspace}%
\newcommand{\effeval}{\texttt{\action\_eff\_eval(\state,\binding)}\xspace}%
\newcommand{\effexec}{\texttt{\action\_eff\_exec(\state,\binding)}\xspace}%
\newcommand{\actparams}{\texttt{\action\_params}\xspace}%

\newcommand{\A}{\mathcal{A}}

\newcommand{\R}{\mathcal{R}}
\newcommand{\E}{\mathcal{E}}
\newcommand{\F}{\mathcal{F}}
\newcommand{\I}{\mathcal{I}}

\newcommand{\schema}{\R}

\definecolor{mgreen}{rgb}{0.0, 0.5, 0.0}

\newcommand{\typename}[1]{\textbf{#1}}
\newcommand{\attrname}[1]{\ensuremath{\mathsf{#1}}}

\newcommand{\scname}[1]{\ensuremath{\mathtt{@#1}}\xspace} % service call name (lowercase)
\newcommand{\actname}[1]{\ensuremath{\mathtt{#1}}\xspace} %
\newcommand{\relname}[1]{\ensuremath{\mathit{#1}}\xspace} % relation name (capitalized)
\newcommand{\cname}[1]{\ensuremath{\mathtt{#1}}\xspace} % constant name
\newcommand{\pname}[1]{{\color{blue!80}\ensuremath{\mathsf{#1}}}\xspace} % param name (lowercase)
\newcommand{\cval}[1]{\mathsf{#1}}

\newcommand{\pl}{\mathcal{P}}

\newcommand{\raw}{\mathtt{raw}}
\newcommand{\slog}{\mathtt{log}}

\newcommand{\dl}{\mathcal{L}}

\tikzstyle{every task} = []
\tikzstyle{every servicetask} = []
\tikzstyle{every gateway} = []
\tikzstyle{every sequence} = []
\tikzstyle{every message} = []
\tikzstyle{every association} = []
\tikzstyle{every event} = []
\tikzstyle{every timer} = []
\tikzstyle{task} = [minimum height=1cm,rectangle,fill=white,thick,draw,rounded corners,align=center, every task, font=\footnotesize]
\tikzstyle{gateway} = [draw,minimum width=7mm, minimum height=8mm,every gateway,fill=white]
\tikzstyle{sequence} = [->,>=triangle 45,every sequence,thick]
\tikzstyle{association} = [->,densely dotted,>=angle 45,every association]
\tikzstyle{event} = [circle,minimum width=7mm, minimum height=7mm,draw,every event,thick,fill=white]
\tikzstyle{start event} = [event, double,fill=white]
\tikzstyle{intermediate event} = [event,double,every event]
\tikzstyle{end event} = [event,line width=2.5pt,fill=white]

\makeatletter

\pgfdeclareshape{XOR}{
  \inheritsavedanchors[from=diamond] % this is nearly a rectangle
  \inheritanchorborder[from=diamond]
  \inheritanchor[from=diamond]{center}
  \inheritanchor[from=diamond]{north}
  \inheritanchor[from=diamond]{south}
  \inheritanchor[from=diamond]{west}
  \inheritanchor[from=diamond]{east}
  \inheritanchor[from=diamond]{north east}
  \inheritanchor[from=diamond]{south east}
  \inheritanchor[from=diamond]{north west}
  \inheritanchor[from=diamond]{south west}

  \backgroundpath{
    \pgf@process{\outernortheast}%
    \pgf@xc=\pgf@x%
    \pgf@yc=\pgf@y%
    \pgfmathsetlength{\pgf@xa}{\pgfkeysvalueof{/pgf/outer xsep}}%
    \pgfmathsetlength{\pgf@ya}{\pgfkeysvalueof{/pgf/outer ysep}}%
    \advance\pgf@xc by-1.414213\pgf@xa%
    \advance\pgf@yc by-1.414213\pgf@ya%
    \pgfpathmoveto{\pgfqpoint{\pgf@xc}{0pt}}%
    \pgfpathlineto{\pgfqpoint{0pt}{\pgf@yc}}%
    \pgfpathlineto{\pgfqpoint{-\pgf@xc}{0pt}}%
    \pgfpathlineto{\pgfqpoint{0pt}{-\pgf@yc}}%
    \pgfpathclose\pgfusepath{stroke}
    \advance\pgf@xc by -.5\pgf@xc%
    \advance\pgf@yc by -.5\pgf@yc%
    \pgfsetlinewidth{2pt}
     \pgfpathmoveto{\pgfqpoint{ 0.8\pgf@xc}{0.8484\pgf@xc}}%
    \pgfpathlineto{\pgfqpoint{-0.8\pgf@xc}{-0.8484\pgf@xc}}%
    \pgfpathmoveto{\pgfqpoint{-0.8\pgf@xc}{0.8484\pgf@xc}}%
    \pgfpathlineto{\pgfqpoint{ 0.8\pgf@xc}{-0.8484\pgf@xc}}%
  }
}

\makeatother

\tikzstyle{table} = [ matrix of nodes,
	ampersand replacement=\&,
    row sep=-\pgflinewidth,
    column sep=-\pgflinewidth,
    nodes={
      rectangle,
      draw=black,
      align=center
    },
    minimum height=1.5em,
    text depth=0.5ex,
    text height=2ex,
    nodes in empty cells,
    every even row/.style={
      nodes={fill=gray!20}
    },
    column 1/.style={
      nodes={font=\ttfamily}
    },
    row 1/.style={
      nodes={
        fill=black,
        text=white,
        font=\bfseries
      }},
      row 3/.style={
      nodes={fill=white!20}},
      row 5/.style={
      nodes={fill=white!20}},
    ]

\tikzstyle{relation}=[rectangle split, rectangle split parts=#1, rectangle split part align=base, draw, anchor=center, align=center, text height=3mm, font=\bfseries, text centered]

\newcolumntype{L}[1]{>{\raggedright\arraybackslash}p{#1}}
\newcolumntype{C}[1]{>{\centering\arraybackslash}p{#1}}
\newcolumntype{R}[1]{>{\raggedleft\arraybackslash}p{#1}}

\usepackage{anyfontsize}
\usepackage{t1enc}
\colorlet{light-gray}{gray!20}
\lstnewenvironment{Java}
{\lstset{
  frame=single,
  backgroundcolor=\color{light-gray},
  basicstyle=\fontsize{8.5}{9}\ttfamily,
  language=Java,
  numbers=left,
  numberstyle=\tiny\color{black},
  captionpos=b,
  keywordstyle=\bfseries\color{green!40!black},
  commentstyle=\itshape\color{purple!40!black},
  identifierstyle=\color{black},
  stringstyle=\color{orange}
}}{}

\definecolor{bronze}{rgb}{0.8, 0.5, 0.2}
\lstnewenvironment{SQL}
{\lstset{
   frameround=ffff,
    language=SQL,    
    numbers=none,
    breaklines=true,
    breakatwhitespace=true,
    mathescape=true,
    tabsize=3,
    keywordstyle=\color{black}\bfseries, 
    morekeywords={VALUES,SET,ENABLES},
    basicstyle=\small\ttfamily\color{black},
}}{}

\newcommand{\sys}{\mathcal{S}}

%% file: abstract.tex
During the last two decades, it has been increasingly acknowledged that the engineering of information systems usually requires a huge effort in integrating master data and business processes. This has led to a plethora of proposals, both from academia and the industry. However, such approaches typically come with ad-hoc abstractions to represent and interact with the data component. This has a twofold disadvantage. On the one hand, they cannot be used to effortlessly enrich an existing relational database with dynamics. On the other hand, they generally do not allow for integrated modelling, verification, and enactment. We attack these two challenges by proposing a declarative approach, fully grounded in SQL, that supports the agile modelling of relational data-aware processes directly on top of relational databases. We show how this approach can be automatically translated into a concrete procedural SQL dialect, executable directly inside any relational database engine. The translation exploits an in-database representation of process states that, in turn, is used to handle, at once, process enactment with or without logging of the executed instances, as well as process verification. The approach has been implemented in a working prototype.

%%% Local Variables:
%%% mode: latex
%%% TeX-master: "main"
%%% End:

%% file: introduction.tex
\section{Introduction}
\label{sec:introduction}

During the last two decades, increasing attention has been given to the
challenging problem, still persisting in modern organizations \cite{Rich10}, of
resolving the dichotomy between business process management and master data
management \cite{Hull08,Reic12,CaDM13}. Devising integrated models and corresponding enactment platforms for processes
and data is now acknowledged as a necessary step to tame a number of conceptual
and enterprise engineering issues, which cannot be tackled by implementation
solutions applied at the level of the enterprise IT infrastructure.

%For
%example, in \cite{MeSW11} it is argued that making business processes
%data-aware is crucial to understand their value, and to evaluate key process
%indicators on top of them. In \cite{Dum11}, Dumas argues that integrating
%data and processes at the modeling level is a key element towards making the business
%logic of such processes explicit, and in turn unambiguously managing business rules without incurring in redundancies and mismatches.

This triggered a flourishing line of research on concrete languages for
data-aware processes, and on the development of tools to model and enact such
processes. The main unifying theme for such approaches is a shift from standard
activity-centric business process meta-models, to a data-centric paradigm that
focuses first on the elicitation of business entities, and then on their
behavioral aspects. Notable approaches in this line are artifact-centric~\cite{Hull08}, object-centric~\cite{KuWR11} and data-centric models~\cite{Reseda18}.
In parallel to these new modeling paradigms, also BPMS based on standard, activity-centric approaches a l\`a BPMN, have
increasingly incorporated data-related aspects in their tool support. Many modern BPM platforms provide (typically proprietary) data models, ad-hoc user interfaces to indicate how
process tasks induce data updates, and query languages to express
decisions based on data. 
%Platforms
%like Bizagi BPM, Bonita BPM, Camunda, and YAWL, all provide (typically
%proprietary) data models, ad-hoc user interfaces to indicate how
%process tasks induce data updates, and query languages to express
%decisions based on data.
%
While this approach has the main advantage of hiding the complexity of the underlying relational database (DB) from the modeler, it comes with two  critical shortcomings.
First, it makes it difficult to conceptually understand the overall process in terms of general, tool-agnostic principles, and to redeploy the same process in a different BPMS. This is witnessed by a number of
ongoing proposals that explicitly bring forward complex mappings for model-to-model transformation (see, e.g., \cite{XSYY11,KopS16}).

Second, this approach cannot be readily applied in the recurrent case where the process needs to be integrated with existing DB tables. In fact, updating an underlying DB through a more abstract data model is an extremely challenging problem that cannot be solved in general, and that is reminiscent to the long-standing, well-known \emph{view update problem} in the database literature \cite{FurC85}. This is often handled by doing the strong assumption that the entire DB schema over which the process (indirectly) operates can be fully generated from the adopted data model, or connected in a \emph{lossless} way to already existing tables. This is, e.g., the approach followed when the process operates over object-oriented data structures that are then linked to an underlying DB via \emph{object-relational mapping techniques}. If existing tables cannot be directly linked to the tool-specific data model, due to an abstraction mismatch between the two layers, then making the process executable on top of such tables requires to guarantee that updates over the data model can be faithfully reproduced in the form of updates on the underlying DB (cf., again, the aforementioned view update problem). This can only be tackled by carefully controlling the forms of allowed updates, by introducing specific data structures acting as a bridge \cite{SuSY16}, and/or by explicitly defining complex mappings to disambiguate how updates should be propagated \cite{SSWY14}.
 
%A second disadvantage comes from the difficulty of incorporating verification capabilities into such systems,which either miss them completely, or only tackle verification of thecontrol-flow perspective \cite{DDG16}. \todo[inline]{Do we need verification?}

In this paper, we address these issues by proposing an alternative approach, called \daphne, where data-aware processes are directly specified on top of standard relational DBs. Our first contribution is a declarative language, called \dapsl, fully grounded in the SQL standard, which allows to:
\begin{inparaenum}[\it (i)]
\item  encapsulate process tasks into SQL-based, parameterized actions that update persistent data possibly injecting values obtained from external inputs (such as user forms, web services, external applications);
\item define rules determining which actions are executable and with which parameter bindings, based on the answers obtained by querying the persistent data. 
\end{inparaenum}
Methodologically, \dapsl can be used either in a bottom-up way as a scripting language that enriches DBs with processes in an agile way, or in a top-down manner as a way to complement standard, control flow-oriented process modelling languages with an unambiguous, runnable specification of conditions and tasks. 
From the formal point of view, \dapsl represents the concrete counterpart of one of the most sophisticated formal models for data-aware processes \cite{BCDDM13}, which comes with a series of (theoretical) results on the conditions under which verification can be carried out. In fact, a wide array of foundational results tackling the formalization of data-ware processes, and the identification of boundaries for their verifiability \cite{CaDM13}, has been produced, but the resulting approaches have never made their way into actual modeling\&enactment tools. In this sense, \dapsl constitutes the first attempt to bridge the gap between such formal approaches and concrete modeling+execution.

Our second contribution is to show how this language is automatically translated into a concrete procedural SQL dialect, in turn providing direct in-database process execution support. This has been implemented within the \daphne-engine, whose back-end consists of a relational storage with corresponding stored procedures to manage the action-induced updates, and whose JAVA front-end provides APIs and functionalities to inspect the current state of the process and its underlying
data, as well as to interact with different concrete systems for acquiring
external data. 

Our third and last contribution is to show that, thanks to a clever way of encoding the process state and related data from \dapsl to SQL, the \daphne engine seamlessly accounts for three key usage modalities: enactment with and without recall about historical data, and state space construction to support formal analysis. Differently from usual approaches in formal verification, where the analysis is conducted on an abstract version of a given concrete model/implementation, \daphne allows one to verify exactly the same process model that is enacted.

%% file: daphne-model.tex
\newcommand{\rds}{RDS\xspace}
\newcommand{\inadom}[1]{\mathit{Live}_{#1}}

%\section{The \daphne Model}\label{sec:dcds}
\section{Data-Aware Process Specification Language}\label{sec:dcds}

%\todo[inline]{Replace DAPS with a language called DAPSL. Mention that this language is essentially a modelling language for (or a practical counterpart of) DCDSs enriched with data types}
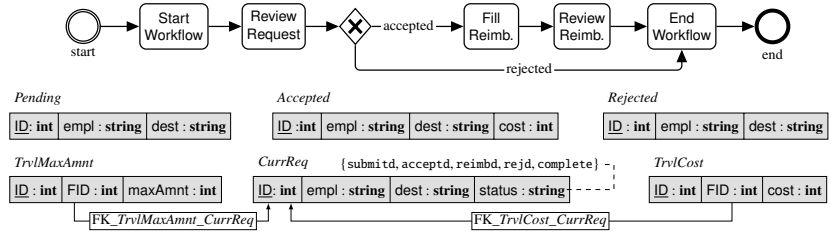
\begin{figure}[t]
\centering
\resizebox{.9\hsize}{!}{

\begin{tikzpicture}
\begin{scope}[auto,node distance=2cm, thick]
%%%%%%%%%%%%%%%%%%%%%%%%%%%%%%%%%%%%%%%%%%%%%%%%%%%%%%%%%%%%%%%%%%%%%%%%
%				PROCESS MODEL
%%%%%%%%%%%%%%%%%%%%%%%%%%%%%%%%%%%%%%%%%%%%%%%%%%%%%%%%%%%%%%%%%%%%%%%%
\node[start event,label=below:start]  (start) at (0,0) {} ;
\node[task] (startw) at (2,0) {\textsf{Start}\\ \textsf{Workflow}} ;
\draw[sequence,->] (start) -- (startw);
\node[task] (revwreq) at (4.2,0) {\textsf{Review}\\ \textsf{Request}} ;
\draw[sequence,->] (startw) -- (revwreq);
\node[XOR,gateway] (gate) at (6,0) {};
\draw[sequence,->] (revwreq) -- (gate);
\node[task] (fillreimb) at (9,0) {\textsf{Fill}\\ \textsf{Reimb.}} ;
\draw[sequence,->] (gate) -- node[fill=white,xshift=-2mm,yshift=-2.2mm,inner sep=.5mm,minimum size=1mm]{\footnotesize{accepted}} (fillreimb);
\node[task] (revwreimb) at (11,0) {\textsf{Review}\\ \textsf{Reimb.}};
\draw[sequence,->] (fillreimb) -- (revwreimb);
\node[task] (endw) at (13.2,0) {\textsf{End}\\ \textsf{Workflow}};
\draw[sequence,->] (revwreimb) -- (endw);
\draw[sequence,->,rounded corners=1mm] (gate.south) -- ++ (0,-0.55) -| node[fill=white,xshift=-28mm,inner sep=.5mm,minimum size=1mm]{\footnotesize{rejected}} (endw.south);
\node[end event,label=below:end]  (end) at (15.2,0) {} ;
\draw[sequence,->] (endw) -- (end);
\end{scope}

%%%%%%%%%%%%%%%%%%%%%%%%%%%%%%%%%%%%%%%%%%%%%%%%%%%%%%%%%%%%%%%%%%%%%%%%
%				DATA MODEL
%%%%%%%%%%%%%%%%%%%%%%%%%%%%%%%%%%%%%%%%%%%%%%%%%%%%%%%%%%%%%%%%%%%%%%%%
\begin{scope}[yshift =-1.6cm,xshift=-.7cm,relation/.style={rectangle split, rectangle split parts=#1, rectangle split part align=base, draw, anchor=center, align=center, text height=3mm, text centered}]\hspace*{-0.3cm}

% RELATIONS

\node (Pending_title) {\relname{Pending}};
\node [relation=3, rectangle split horizontal, rectangle split part fill={lightgray!50}, below=0.6cm of Pending_title.west, anchor=west] (Pending)
{\underline{\attrname{ID}}: \typename{int} %
\nodepart{two} \attrname{empl} : \typename{string}
\nodepart{three} \attrname{dest} : \typename{string}};

\node [right=5.8cm of Pending_title.west, anchor=west] (Accepted_title) {\relname{Accepted}};
\node [relation=4, rectangle split horizontal, rectangle split part fill={lightgray!50}, anchor=north west, below=0.6cm of Accepted_title.west, anchor=west] (Accepted)
{\underline{\attrname{ID}} :\typename{int}%
\nodepart{two} \attrname{empl} : \typename{string}
\nodepart{three}  \attrname{dest} : \typename{string}
\nodepart{four} \attrname{cost} : \typename{int}};

\node [right=5.9cm of Accepted_title.east, anchor=west] (Rejected_title) {\relname{Rejected}};
\node [relation=3, rectangle split horizontal, rectangle split part fill={lightgray!50}, anchor=north west, below=0.6cm of Rejected_title.west, anchor=west] (Rejected)
{\underline{\attrname{ID}} : \typename{int}%
\nodepart{two} \attrname{empl} : \typename{string}
\nodepart{three}  \attrname{dest} : \typename{string}};

\node [xshift=-.0cm,below=0.8cm of Pending.west, anchor=west] (TrvlMaxAmnt_title) {\relname{TrvlMaxAmnt}};
\node [relation=3, rectangle split horizontal, rectangle split part fill={lightgray!50}, anchor=north west, below=0.6cm of TrvlMaxAmnt_title.west, anchor=west] (TrvlMaxAmnt)
{\underline{\attrname{ID}} : \typename{int}%
\nodepart{two}  \attrname{FID} : \typename{int}
\nodepart{three}  \attrname{maxAmnt} : \typename{int}};

\node  [right=3.4cm of TrvlMaxAmnt_title.east, anchor=west](CurrReq_title) {\relname{CurrReq}};
\node [relation=4, rectangle split horizontal, rectangle split part fill={lightgray!50}, anchor=north west, below=0.6cm of CurrReq_title.west, anchor=west] (CurrReq)
{\underline{\attrname{ID}}: \typename{int}%
\nodepart{two} \attrname{empl} : \typename{string}
\nodepart{three}  \attrname{dest} : \typename{string}
\nodepart{four} \attrname{status} : \typename{string}};

\node[above=0.6cm of CurrReq.south,xshift=1.29cm] (values) {$\set{\cname{submitd},\cname{acceptd},\cname{reimbd},\cname{rejd},\cname{complete}}$};

\node [right=7.4cm of CurrReq_title.east, anchor=west] (TrvlCost_title) {\relname{TrvlCost}};
\node [relation=3, rectangle split horizontal, rectangle split part fill={lightgray!50}, anchor=north west, below=0.6cm of TrvlCost_title.west, anchor=west] (TrvlCost)
{\underline{\attrname{ID}} : \typename{int}%
\nodepart{two}  \attrname{FID} : \typename{int}
\nodepart{three}  \attrname{cost} : \typename{int}};

% FOREIGN KEYS + CHECK CONSTRAINTS

\draw[-latex] ($(TrvlMaxAmnt.two south) +(-0.4,0)$) -- ++ (0,-0.40) -| node[rectangle,draw=black,fill=white,xshift=-21mm,inner sep=.5mm,minimum size=1mm]{\footnotesize{\textup{FK\_\relname{TrvlMaxAmnt}\_\relname{CurrReq}}}} ($(CurrReq.one south) + (-0.2,0)$);
\draw[-latex] (TrvlCost.two south) -- ++ (0,-0.40) -| node[rectangle,draw=black,fill=white,xshift=55mm,inner sep=.5mm,minimum size=1mm]{\footnotesize{\textup{FK\_\relname{TrvlCost}\_\relname{CurrReq}}}} ($(CurrReq.one south) + (0.3,0)$);
%\draw[dashed] (values.north) -- (CurrReq.four south);
\draw[dashed] (values.east) --++  (0,0) -| ($(values.east) + (0.2,0)$) |- ($(CurrReq.east) + (-0.1,0)$) ;

\end{scope}
\end{tikzpicture}
}
\caption{The travel management process and a corresponding data model}
\label{fig:ex-bpm}
\end{figure}

\daphne relies on a declarative, SQL-based \emph{data-aware processes specification language} (\dapsl) to capture processes operating over relational data. \dapsl provides a SQL-based front-end conceptualization of \emph{data-centric dynamic systems} (DCDSs)~\cite{BCDDM13}, one of the most well-known formal models for data-aware processes. Methodologically, \dapsl can be seen as a guideline for business process programmers that  have minimal knowledge of SQL and aim at developing process-aware, data-intensive applications.

A \dapsl specification consists of two main components: 
\begin{inparaenum}[\it (i)]
\item a \emph{data layer}, which accounts for the structural aspects of the domain, and maintains its corresponding extensional data;
\item a \emph{control layer}, which inspects and evolves the (extensional part of the) data layer.
\end{inparaenum}
%\daphne relies on a declarative, SQL-based \emph{data-aware process specification} (\dapm) to capture processes operating over relational data. Methodologically, \dapm can be seen as a guideline for business process programmers that  have minimal knowledge of SQL and aim at developing process-aware, data-intensive applications. Technically, \dapm can be seen as a SQL-based variant of \emph{data-centric dynamic systems} (DCDS) \cite{BCDDM13}, one of the most well-known formal models for data-aware processes. 
%x%
%\subsection{Data and Control Layers}
%A \dapm consists of two components:
%\begin{inparaenum}[\it (i)] 
%\item a \emph{data layer}, which accounts for the structural aspects of the domain, and maintains its corresponding extensional data;
%\item a \emph{control layer}, which inspects and evolves the (extensional part of the) data layer.
%\end{inparaenum}
%
We next delve into these two components in detail, illustrating
the essential features of \dapsl on the following running example, inspired by \cite{DDG16}.

\begin{example}
  \label{exa:rds}
  We consider a travel reimbursement process, whose control flow is depicted in
  Figure~\ref{fig:ex-bpm}.
  %
  %% The high level representation of its control flow in
  %% Figure~\ref{fig:ex-bpm} captures the organization of action calls over a
  %% database used to manage travel expenses.
  The process starts (\textsf{StartWorkflow}) by checking pending employee
  travel requests in the database. Then, after selecting a request, the system
  examines it (\textsf{ReviewRequest}), and decides whether to approve it or
  not. If approved, the process continues by calculating the maximum
  refundable amount, and the employee can go on her business trip. On arrival,
  she is asked to compile and submit a form with all the business trip expenses
  (\textsf{FillReimb}). The system analyzes the submitted form
  (\textsf{ReviewReimb}) and, if the estimated maximum has not been exceeded,
  approves the refunding. Otherwise the reimbursement is rejected.
  \qedtriangle
\end{example}

%\subsection{Formal Model}

%By design, the two layers interact as follows: the data layer stores all
%the data of interest inside a full-fledged relational database with constraints, 
%while the process layer modifies and evolves such data, possibly injecting (fresh) data 
%from the external environment. 

%A DCDS is a tuple $\dcds=\tup{\dlayer,\player}$, where $\dlayer$ is called the
%\emph{data layer} and $\player$ the \emph{process layer} of $\dcds$.  The two
%layers interact as follows: the data layer stores all the data of interest
%inside a full-fledged relational database with constraints, while the process
%layer modifies and evolves such data, possibly injecting fresh data.
%x\smallskip
\noindent
\textbf{Data layer.} Essentially, the \emph{data layer} is a standard relational DB, consisting of an intensional (schema) part, and an extensional (instance) part. The intensional part is a \emph{database schema}, that is, a pair $\tup{\schema,\E}$, where $\schema$ is a finite set of \emph{relation schemas}, and $\E$ is a finite set of \emph{integrity constraints} over $\schema$. To capture a database schema, \dapsl employs the standard SQL data definition language (DDL). For presentation reasons, in the remainder of the papers we refer to the components of a database schema in an abstract way, following standard definitions recalled next. As for relation schemas, we adopt the \emph{named perspective}: a relation schema is defined by a signature containing a \emph{relation name} and a set of \emph{attribute names}.
We implicitly assume that all attribute names are typed, and so are all the constitutive elements of a \dapsl model that insist on relation schemas. Type compatibility can be easily defined and checked, again thanks to the fact that \dapsl employs the standard SQL DDL. 

\dapsl tackles three fundamental types of integrity constraints: \emph{primary keys}, \emph{foreign keys}, and \emph{domain constraints}. A domain constraint is attached to a given relation attribute, and explicitly enumerates which values can be assigned to that attribute. 
    To succinctly refer to primary and foreign keys, we use the following shorthand notation. Given  a relation $\relname{R}$ and a tuple $A=\tup{\attrname{A}_1,\ldots, \attrname{A}_m}$ of attributes defined on $\relname{R}$, we use  $\proj{\relname{R}}{A}$ to denote the projection of $\relname{R}$ on such attributes. $\pk{R}$ indicates the set of attributes in $R$ that forms its primary key (similarly for keys), whereas $\FK{\relname{S}}{B}{\relname{R}}{A}$ defines that projection $\proj{\relname{S}}{B}$ is a foreign key referencing $\proj{\relname{R}}{A}$ (where attributes in $B$ and those in $A$ are matched component-wise). Recall, again, that such constraints are expressed in \dapsl by using that standard SQL DDL.
   
The extensional part of the data layer is a \emph{DB instance} (which we simply call DB for short). \dapsl delegates the representation of this part to the relational storage of choice. Abstractly, a database  consists of a set of labelled tuples over the relation schemas in $\schema$. Given a relation $R$ in $\schema$, a labelled tuple over $R$ is a total function mapping the attribute names in the signature of $R$ to corresponding values. We always assume that a database is consistent, that is, satisfies all constraints in $\E$. While the intensional part is fixed in a \dapsl model, the extensional part starts from an initial database that is then iteratively updated through the control layer, as dictated below.

    %We assume type consistency in the definition of foreign keys and domain constraints. 
    
% The \emph{data layer} structures and maintains the relevant
%data about the domain of interest. For that, we adopt the most generic relational 
%database representation, that is, a set of relation names $\schema$ and a set of constraints $\E$ defined on them. In what follows, we employ the following notational conventions. 
%Here, a relational schema $\schema$ is a finite set of relations $\relname{R}(A)$,  where $\relname{R}$ is a relation name and $A$ is an ordered set of attributes $\tup{\attrname{A}_1,\ldots, \attrname{A}_m}$ defined on $\relname{R}$, and  $\proj{\relname{R}}{A}$ denotes the projection of $\relname{R}$ on such attributes.
%In turn, $\E$ is a set of integrity constraints defined on $\schema$.  In our model we consider two types of integrity constraints: primary keys and foreign keys. Here, 
%\todo[inline]{mention how to check validity of domain constraints (use an ad-hoc way within $\alpha$\_eff\_exec actions).}

\begin{example}
  \label{exa:rds-data-layer}
  The \dapsl DB schema  for the process informally described in Example~\ref{exa:rds} is shown in Figure~\ref{fig:ex-bpm}.
  % As one can see, each relational schema is defined by its name and a list of
  % typed attributes.
  We recall of the relation schemas:
  % of the schema in Figure~\ref{fig:ex-bpm}:
  \begin{inparaenum}[\itshape (i)]
  \item requests under process are stored in the relation
    % defined by a schema
    $\relname{CurrReq}$,
    % $\relname{CurrReq}(\attrname{ID}:\typename{int},\attrname{empl}:\typename{string},
    % \attrname{dest}:\typename{string},\attrname{status}:\typename{string})$,
    whose components are the request UID, which is the primary key,
    % (i.e., $\PK{CurrReq}{\set{\attrname{ID}}}$),
    the employee requesting a reimbursement, the trip destination, and the
    status of the request, which ranges over a set of predefined values (captured with a domain constraint);
  \item maximum allowed trip budgets are stored in $\relname{TrvlMaxAmnt}$,
    whose components are the id (the primary key),
    % (i.e., $\PK{TrvlMaxAmnt}{\set{\attrname{ID}}}$),
    the request reference number (a foreign key),
    % (i.e.,
    % $\FK{TrvlMaxAmnt}{CurrReq}{\set{\attrname{FID}}}{\set{\attrname{ID}}}$),
    and the maximum amount assigned for the trip;
  \item $\relname{TrvlCost}$ stores the total amount spent, with the same
    attributes as in $\relname{TrvlMaxAmnt}$.
  \end{inparaenum}
  \qedtriangle
\end{example}

%The \emph{process layer} constitutes the progression mechanism for the DCDS. It
%is represented as a tuple $\player=\tup{\services,\A,\rho}$, where $\services$
%is a finite set of (uninterpreted) functions representing calls to external
%services, $\A$ is a finite set of (update) actions, and $\rho$ is a process
%specification.

\noindent
\textbf{Control layer.} The \emph{control layer} defines how the data layer can be evolved through the execution of actions (concretely accounting for the different process tasks). Technically, the
control layer is a triple $\tup{\F,\A,\rho}$, where $\F$ is a finite set of \emph{external services}, $\A$ is a finite set of \emph{atomic tasks} (or actions), and $\rho$ is a \emph{process specification}. 
%Similarly to the data layer, the definition of the control layer coincides with the one of DCDSs~\cite{BCDDM13}.s

 Each service is a described as a function signature that indicates how externally generated data can be brought into the process, abstractly accounting for a variety of concrete data injection mechanisms such as user forms, third-party applications, web services, internal generation of primary keys, and so on. Each external service comes with a signature indicating the \emph{service name}, its \emph{formal input parameters} and their \emph{types}, as well as the \emph{output type}.

Actions are the basic building blocks of the control layer, and represent transactional operations over the data layer.  
Each action comes with a distinguished name and a set of formal parameters, and consists of a set of \emph{parameterized SQL statements} that inspect and update the current state of the \dapsl model (i.e., the current DB), using standard insert-delete SQL operations. Such operations are parameterized so as to allow referring with the statements to the action parameters, as well as to the results obtained by invoking a service call. Both kind of parameters are substituted with actual values when the action is concretely executed.
Hence, whenever a SQL statement allows for using a constant value, \dapsl allows for using either a constant, an action parameter, or a placeholder representing the invocation of a service call. To clearly distinguish service invocations from action parameters, \dapsl prefixes the service call name with symbol $\scname{}$.

%
%. We 
%
% Some actions can also
%contain external service calls (from $\F$) that, upon invocation, provide data
%possibly not present in the current database. Every action execution guarantees
%that all the respective updates produce a database instance satisfying all the
%constraints in $\E$.  If this is not the case, all the action updates are
%rolled back.

Formally, an \dapsl \emph{action} is an expression $\alpha(p_1,\ldots,p_n){\,:\,}\{e_1;\ldots;e_m\}$, where:
\begin{compactitem}
\item $\alpha(p_1,\ldots,p_n)$ is the action \emph{signature}, constituted by   
\emph{action name} $\alpha$ and the set $\set{p_1,\ldots,p_n}$ of \emph{action formal parameters};
\item $\{e_1,\ldots,e_m\}$ is a set of parameterized SQL insertions and deletions.
\end{compactitem}
We assume that no two actions in $\A$ share the same name, and then use the action name to refer to its corresponding  specification. 
Each effect specification $e_i$ modifies the current DB using standard SQL, and is either a deletion or insertion.  

A \emph{parameterized SQL insertion} is a SQL statement of the form:
\begin{SQL}
		INSERT INTO $\relname{R}(\attrname{A}_1,\ldots, \attrname{A}_k)$ VALUES $(t_1,\ldots,t_k),$
\end{SQL} 
where $R$ is the name of relation schema, and each $t_j$ is either a value, an action formal parameter or a service call invocation (which in SQL syntactically corresponds to a scalar function invocation). Given a service call $\scname{F}$ with $p$ parameters, an invocation for $F$ is of the form $\scname{F}(x_1,\ldots,x_p)$, where each $x_j$ is either a value or an action formal parameter (i.e., functions are not nested). 
%%
%The invocation intuitively denotes that when the insert effect is actually applied by instantiating the action formal parameters with specific values,  
Notice that \lstinline{VALUES}{} can be seamlessly substituted by a complex SQL selection inner query, which in turn allows for \emph{bulk insertions} into $\relname{R}$, using all the answers obtained from the evaluation of the inner query.

%\item \verb{INSERT INTO{ $\relname{R}(\attrname{A}_1,\ldots, \attrname{A}_k)$ $expr(t_1,\ldots,t_k)$,\todo{shall we use \texttt{expr}$(t_1,\ldots,t_k)$ or  $\tup{expr}$?} where an expression $expr$ is either a table value constructor \verb{VALUES{ or a \verb{SELECT{ query,\todo{since we don't support default values, it's important to mention explicitly all the relational arguments in expr.}  and each $t_j$ can be either a value, an action parameter or a scalar function. In turn, every scalar function, defined as $\mathtt{@F}(x_1,\ldots,x_p)$, is used to represent external service calls, and model the values that these calls return from the external environment while being invoked during action execution. We assume all the scalar functions to be flat, that is, every $x_j$ can only be a value or an action parameter.

A \emph{parameterized SQL deletion} is a SQL statement of the form:
\begin{SQL} 
		DELETE FROM $\relname{R}$ WHERE $\tup{condition},$
\end{SQL}
 where $R$ is the name of relation schema, and the \lstinline{WHERE}{} SQL clause may internally refer to the  action formal parameters. This specification captures the simultaneous deletion of all tuples returned by the evaluation of $condition$ on the current DB. 
 Consistently with classical conceptual modeling approaches to domain changes \cite{Oli07}, we allow for overlapping deletion and insertion effect specifications, giving higher priority to deletions, that is, first all deletions and then all insertions are applied. This, in turn, allows to unambiguously capture update effects (by deleting certain tuples, and inserting back variants of those tuples). Introducing explicit SQL update statements would in fact create ambiguities on how to prioritize updates w.r.t.~potentially overlapping deletions and insertions.
%%
%
%
%
%. Introducing a ``classical'' SQL updates can be seamlessly performed, but may cause semantical problems in cases when, within an \lstinline{UPDATE}{} statement, the process modeller accidentally introduces data modifications overlapping with other action effects that contain \lstinline{DELETE}{}s and \lstinline{INSERT}{}s.

%\begin{SQL}
%	(3) UPDATE $\relname{R}$ SET $\tup{new\textrm{-}value\: assignments}$ WHERE $\tup{condition},$ 
%\end{SQL}
% where $\tup{new\textrm{-}value\: assignments}$, similarly to \lstinline{INSERT}{} effects, can be parametrized either with action parameters or service calls, while $\tup{condition}$ can only include action parameters in its expressions. 
%\todo[inline]{mention ISO SQL standard}

The executability of an action, including how its formal parameters may be bound to corresponding values, is dictated by the process specification $\rho$ --  a set of condition-action (CA) rules, again grounded in SQL, and used to declaratively capture the control-flow of the \dapsl model.  

For each action $\alpha$ in $\A$ with $k$ parameters, $\rho$ contains a single CA rule determining the executability of $\alpha$. The CA rule is an expression of the form:
\begin{SQL}
		SELECT $\attrname{A}_1,\ldots, \attrname{A}_s$ FROM $\relname{R_1},\ldots, \relname{R_m}$ WHERE $\tup{condition}$ 
	     	ENABLES $\alpha(\attrname{A}_{n_1},\ldots,\attrname{A}_{n_k}),$
\end{SQL}
where each $\attrname{A}_i$ is an attribute, each $\relname{R}_i$ is the name of a relation schema, $\alpha\in\A$, and $\set{\attrname{A}_{n_1},\ldots,\attrname{A}_{n_k}}\subseteq\set{\attrname{A}_1,\ldots, \attrname{A}_s}$.
Here, the SQL \lstinline{SELECT} query represents the rule condition, and the results of the query provide alternative actual parameters that instantiate the formal parameters of $\alpha$. This grounding mechanism is applied on a per-answer basis, that is, to execute $\alpha$ one has to choose how to instantiate the formal parameters of $\alpha$ with one of the query answers returned by the \lstinline{SELECT} query. Multiple answers consequently provide alternative instantiation choices.
%This allows on to specify and execute data-aware processes with case-specific, parameterizable activities. W.l.o.g., we assume that there is a single CA rule for each action.
Notice that requiring each action to have only one CA rule is without loss of generality, as multiple CA rules for the same action can be compacted into a unique rule whose condition is the \lstinline{UNION}{} of the condition queries in the original rules.

\begin{example}
  \label{exa:rds-process-layer}
  % To define the process of the DCDS in Example~\ref{exa:rds}, we introduce
  % three uninterpreted functions modeling service calls: $\scname{status}$
  % updates the request status by taking as input an employee name and a trip
  % destination, and returning a status value; $\scname{maxAmnt}$ returns the
  % maximum refundable amount assigned to accepted requests, taking as input
  % the same arguments as $\scname{status}$; and $\scname{cost}$ returns the
  % travel cost obtained through a user form, and has the same arguments as
  % $\scname{maxAmnt}$.
  We concentrate on three tasks of the process in
  Example~\ref{exa:rds}, and show their \dapsl representation.
  Task \textsf{StartWorkflow} creates a new travel reimbursement request by picking one of the pending requests from the current database. We represent this in \dapsl as an action with three formal parameters, respectively denoting a pending request id, its responsible employee, and her intended destination:
%  \textsf{StartWorkflow} is modeled as action $\actname{StartWorkflow}$, with
%  the precondition
  \begin{SQL}
SELECT $\attrname{id},\attrname{empl},\attrname{dest}$ FROM $\relname{Pending}$ ENABLES $\actname{StartWorkflow}(\attrname{id},\attrname{empl},\attrname{dest})$

$\actname{StartWorkflow}(\pname{id},\pname{empl},\pname{dest})$:$\big\{$ DELETE FROM $\relname{Pending}$ WHERE $\relname{Pending}.\attrname{id} = \pname{id}$;
$\phantom{\actname{StartWorkflow}(\pname{id},\pname{empl},\pname{dest})\text{:}\big\{}$ INSERT INTO $\relname{CurrReq}(\attrname{id},\attrname{empl},\attrname{dest},\attrname{status})$
$\phantom{\actname{StartWorkflow}(\pname{id},\pname{empl},\pname{dest})\text{:}\big\{}\quad$ VALUES$(\scname{genpk}(),\pname{empl},\pname{dest},\cname{submitd})~\big\}$
  \end{SQL}
 Here, a new travel reimbursement request is generated by removing the entry of $\relname{Pending}$ matching the given $\pname{id}$, and then inserting a new tuple into $\relname{CurrReq}$ by passing the $\attrname{empl}$ and $\attrname{dest}$ values of the deleted tuple, and by setting the status to $\cname{\color{bronze}\textsf{`}{submitd}\textsf{'}}$. To get a unique identifier value for the newly inserted tuple, we invoke the nullary service call $\scname{genpk}$, which returns a fresh primary key value. This can be avoided in practice given that the corresponding primary key field generates unique values automatically. 
% \todo[inline]{explain why to use nullary service and that if the CurrReq table has the first attribute defined as a primary key in a RDBMS, then the primary key attribute value will be automatically incremented}
 % 
  %
  
  Task \textsf{ReviewRequest} examines an employee trip request and,
  if accepted, assigns the maximum reimbursable amount to it.  The corresponding action can be
  executed only if the $\relname{CurrReq}$ table contains at least one submitted request:
  \begin{SQL}
  SELECT $\attrname{id},\attrname{empl},\attrname{dest}$ FROM $\relname{CurrReq}$ WHERE $\relname{CurrReq}.\attrname{status}=\cname{\color{bronze}\textsf{`}{submitd}\textsf{'}}$ 
  $\quad$ ENABLES $\actname{RvwRequest}(\attrname{id},\attrname{empl},\attrname{dest})$

  $\actname{RvwRequest}(\pname{id},\pname{empl},\pname{dest})$:$\big\{$DELETE FROM $\relname{CurrReq}$ WHERE $\relname{CurrReq}.\attrname{id} = \pname{id}$;
   $\qquad\qquad\qquad\qquad\qquad\qquad$INSERT INTO $\relname{CurrReq}(\attrname{id},\attrname{empl},\attrname{dest},\attrname{status})$
  $\qquad\qquad\qquad\qquad\qquad\qquad\qquad$VALUES$(\pname{id},\pname{empl},\pname{dest},\scname{status}(\pname{empl},\pname{dest}))\big\}$;
  	$\qquad\qquad\qquad\qquad\qquad\qquad$INSERT INTO $\relname{TrvlMaxAmnt}(\attrname{tid},\attrname{tfid},\attrname{tmaxAmnt})$
  	$\qquad\qquad\qquad\qquad\qquad\qquad\qquad$VALUES$(\scname{genpk}(),\pname{id},\scname{maxAmnt}(\pname{empl},\pname{dest}))\big\}$
  \end{SQL}  
% $\phantom{xxxxxxxxxxxxxxxxxxxxxx}$        AND $\relname{CurrReq}.\attrname{status} = \cname{submitted}$; 
%  \begin{SQL}
%  $\actname{RvwRequest}(\pname{id},\pname{empl},\pname{dest})$:$\big\{$UPDATE $\relname{CurrReq}$ 
%  $\phantom{xxxxxxxxxxxxxxxxxxx}$SET $CurrReq.\attrname{status} = \mathtt{@status}(\pname{empl},\pname{dest})$ 
% 	$\phantom{xxxxxxxxxxxxxxxxxx}$WHERE $\relname{CurrReq}.\attrname{id}=\pname{id}$;
%  	$\phantom{xxxxxxxxxxxxxxxxxx}$INSERT INTO $\relname{TrvlMaxAmnt}(\attrname{id},\attrname{fid},\attrname{maxAmnt})$
%  	$\phantom{xxxxxxxxxxxxxxxxxx}$VALUES$(\scname{genpk}(),\pname{id},\scname{maxAmnt}(\pname{empl},\pname{dest}))\big\}$
%  \end{SQL}
%  $\begin{array}{ll}
%\actname{RvwRequest}(\pname{id},\pname{empl},\pname{dest}) : \big\{&\texttt{UPDATE } \relname{CurrReq} \\
%&\texttt{SET } CurrReq.\attrname{status} = \mathtt{@status}(\pname{empl},\pname{dest})\\
%&\texttt{WHERE } \relname{CurrReq}.\attrname{id}=\pname{id};\\
%&\texttt{INSERT INTO }\relname{TrvlMaxAmnt}(\attrname{id},\attrname{fid},\attrname{maxAmnt})\\
%&\texttt{VALUES}(\scname{genpk}(),\pname{id},\scname{maxAmnt}(\pname{empl},\pname{dest}))\big\}
%  \end{array}$
The request status of $\relname{CurrReq}$ is updated by calling service  $\scname{status}$, that takes as input an employee name and a trip destination, and returns a new status value. 
  Also, a new tuple containing the maximum reimbursable amount is added to $\relname{TrvlMaxAmnt}$.  To get the maximum refundable amount  for $\relname{TrvlMaxAmnt}$, we employ service $\scname{maxAmnt}$ with the same arguments as $\scname{status}$.

  Task \textsf{FillReimbursement} updates the current
  request by adding a compiled form with all the trip expenses. This
  can be done only when the request has been accepted:
  \begin{SQL}
  SELECT $\attrname{id},\attrname{empl},\attrname{dest}$ FROM $\relname{CurrReq}$ WHERE $\relname{CurrReq}.\attrname{status}=\cname{\color{bronze}\textsf{`}{acceptd}\textsf{'}}$  
  $\qquad$ENABLES $\actname{FillReimb}(\attrname{id},\attrname{empl},\attrname{dest})$
  
  $\actname{FillReimb}(\pname{id},\pname{empl},\pname{dest})$:$\big\{$INSERT INTO $\relname{TrvlCost}(\attrname{id},\attrname{fid},\attrname{cost})$
$\qquad\qquad\qquad\qquad\qquad\qquad\qquad$VALUES$(\scname{genpk}(),\pname{id},\scname{cost}(\pname{empl},\pname{dest}))\big\}$
  \end{SQL}
  Again, $\scname{genpk}$ and $\scname{cost}$ are used to 
  obtain values externally upon insertion.
  \qedtriangle
\end{example}

\noindent
\textbf{Execution semantics}. First of all, we define the execution semantics of \dapsl actions. Let $\I$ be the current database for the data layer of the \dapsl model of interest. 
An action $\alpha$ is \emph{enabled} in $\I$ if the evaluation of the SQL query constituting the condition in the CA rule of $\alpha$ returns a nonempty result set. This result set is then used to instantiate $\alpha$, by non-deterministically picking an answer tuple, and use it to bind the formal parameters of $\alpha$ to actual values. This produces a so-called \emph{ground action} for $\alpha$. 
The execution of a ground action amounts to simultaneous application of all its effect specifications, which requires to first manage the service call invocations, and then apply the deletions and insertions.  This is done as follows. First, invocations in the ground action are instantiated by resolving the subqueries present in all those 
insertion effects whose values contain invocation placeholders. Each  invocation then becomes a fully specified call to the corresponding service, passing the ground values as input. The result obtained from the call is used to replace the invocation itself, getting a fully instantiated \lstinline{VALUES}{} clause for the insertion effect specification.
%In turn, invoked service calls provide ``missing'' values for completing desired database 
%modifications. Secondly, even though it may seem that every action represents a homogeneous SQL script composed out of effect queries, an execution 
%order has to be enforced as to avoid generation of incorrect (both w.r.t. the business 
%logic and integrity constraints) database instances. Thus, we stipulate all the action \lstinline{DELETE}s{} to precede the action \lstinline{INSERT}s{}, and require  both operation blocks to be executed simultaneously.
Once this instantiation is in place, a transactional update is issued on $\I$, first pushing all deletions, and then  all insertions. If the resulting DB satisfies the constraints of the data layer, then the update is committed. If instead some constraint is violated, then the update is rolled back, maintaining $\I$ unaltered. The explained semantics fully mimics the formal execution semantics DCDSs~\cite{BCDDM13}.

\smallskip
\noindent \textbf{Connection with existing process modeling languages.} In the introduction, we mentioned that \dapsl can be used either as a declarative, scripting language to enrich a DB with process dynamics, or to complement a control-flow process modeling language with the concrete specification of conditions and tasks. The latter setting however requires a mechanism to fully encode the resulting process into \dapsl itself. Thanks to the fact that \dapsl represents a concrete counterpart for DCDSs, we can take advantage from the quite extensive literature showing how to encode different process modeling languages into DCDSs. 
%
%As pointed out before, \dapsl is virtually a SQL-based variant of the DCDS formal framework. This has a twofold advantage. On the one hand, all the results on the boundaries of verifiability for DCDS directly transfer to \dapm as well. On the other hand, DCDS have been used to formalize a variety of more concrete process modeling languages, consequently showing that a declarative, rule-based mechanism for defining the process control-flow can be used to encode both data- and activity-centric approaches. 
In particular, \dapsl can be readily used to capture:
%Previously conducted research showed that DCDS models, thanks to their expressive power, can 
%be automatically constructed from other input specification languages, such as:
%This is not an issue in our setting. 
%In fact, previous research has investigated how DCDS models can be automatically constructed 
%from other input specification languages, such as:
\begin{inparaenum}[\it (i)]
\item data-centric process models supporting the explicit notion of process instance (i.e., case) \cite{MonC16};
\item several variants of Petri nets \cite{BCDM14} without data;
\item Petri nets equipped with resources and data-carrying tokens \cite{MonR16};
\item recent variants of (colored) Petri nets equipped with a DB storage \cite{DDG16,MonR17,RMRS18};
\item artifact-centric process models specified using Guard-Stage-Milestone language  by IBM \cite{SMTD13}.
\end{inparaenum}
The translation rules defined in these papers can be readily transformed into model-to-model transformation rules using \dapsl as target.

%% file: daphne-framework.tex
% Relations in DAPHNE acquire a multi-layered structure, where each relation
% $R$, from the set of relational schemas of the underlying DCDS model, has its
% persistence data storage and representation components separated.

\section{The \daphne System}
\label{sec:daphne}

We discuss how \dapsl has been implemented in a concrete system that provides in-database process enactment, as well as the basis for formal analysis.

%%The language presented in Section~\ref{sec:dcds} has been
%%implemented in a system called \daphne. The system
%%provides functionalities for modelling and execution of data-aware
%%processes. \todo{enriched with service calls?}
%%%%
%%In this section we introduce \daphne by presenting its
%%architecture and providing details about the implementation of
%%$\dapl$. 

%
%The system
%provides two main functionalities: DCDS execution and
%DCDS abstraction. It should be noted that the former represents the core functionality of the
%system, while the latter comes as an extension on top of it.
%%%
%In this section we introduce \daphne by presenting its
%architecture and providing details about the implementation of
%its two functionalities. For simplicity, we first focus on execution
%and then abstraction.

\begin{figure}[tb]
  \centering
  \includegraphics[width=.85\textwidth]{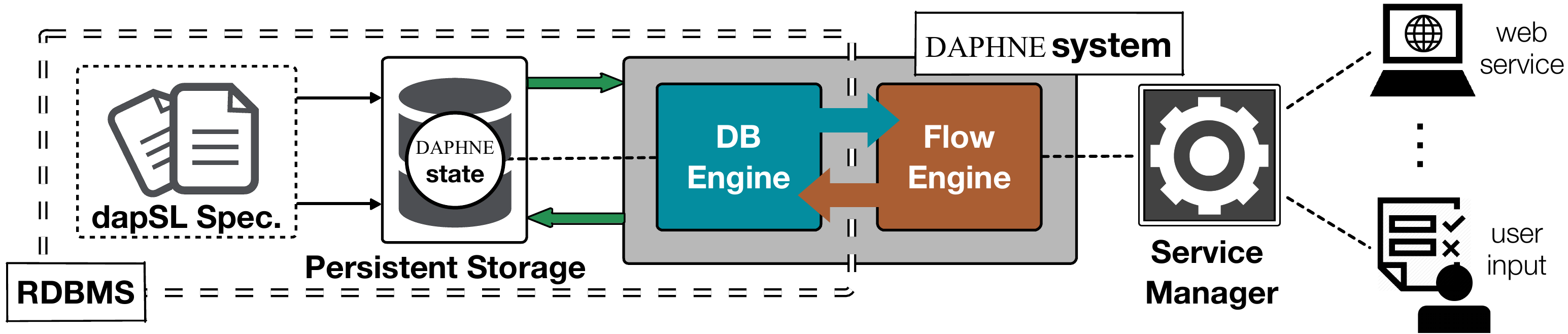}
  \caption{Conceptual architecture of \daphne%; a \daphne specification is a concrete representation of a \dapm   
  %\todo[inline]{refine the schema: add detailed DCDS specification using the DAPHNE language processor/translator}
}
  \label{fig:architecture}
\end{figure}

\subsection{Internal Architecture}
\label{sec:architecture}

The core architecture of \daphne is depicted in Figure~\ref{fig:architecture}.
The system takes as input a representation of a \dapsl specification (or model)  and uses a standard DBMS to support its execution.

The DBMS takes care of storing the data relevant to the input \dapsl model and supports,
through the \emph{\dbengine} of the underlying DBMS, the application of a set of operations that jointly realize the given \dapsl actions.
The \emph{\fengine} constitutes the application layer of the system;
it facilitates the execution of a \dapsl model by coordinating the
activities that involve the user, the DBMS, and the services.
Specifically, the \fengine issues queries to the DBMS, calls stored procedures,
and handles the communication with external services through a further module called \emph{\smanager}.

Next we give a detailed representation of \daphne's architecture by describing
the stages of each execution step. For the moment we do not consider how the input \dapsl specification is concretely encoded inside the DBMS.  
At each point in time, the DBMS stores the
current state of the \dapsl model. We assume that, before the execution starts, the DBMS contains an initial database instance for the data layer of \dapsl model.
To start the
execution, %As a first step,
the \fengine queries the DBMS about the actions that are enabled in the current state; if one is found, the engine retrieves all possible parameter assignments that can be selected to ground the action, and returns them to the user (or the software module responsible for the process enactment). The user is then asked to choose one of such parameter assignments.
% and presents them to the user. The user then selects the parameters over which the action is going to be executed.
%%
At this point, the actual application of the ground action is triggered. The \fengine invokes a set of stored procedures from the DBMS that
 take care of evaluating and applying action effects.
If needed by the action specification, the \fengine interacts with
external services, through the \smanager, to acquire new data
via service calls. The tuples to be deleted and inserted in the various relations of the \dapsl model are then computed, and the consequent changes are pushed to the DBMS within a transaction, so that the underlying database instance is updated only if all constraints are satisfied. 
%
%The state resulting from the action execution is committed to the database through a transaction, which is aborted if some constraint specified
%in the data layer is violated.
%%
After the update is committed or rolled back, the action execution cycle can be repeated by selecting either a new parameter assignment or another action available in the newly generated state.

%We first observe that when presenting basic execution
%above, we have considered only the current state stored in
%the DBMS.
%%%\todo[inline]{However, before that we would like to note that the basic execution described above is stateless: the system maintains only the most recently updated state.
%%%{\bf FP: NO. it's stateful: we record the current state!}}
%\todo[inline]{Justify the historical storage}
%In fact, however, \daphne is designed to store a full history
%and even the whole (abstraction of the) transition system associated
%with the DCDS.
%%%\todo[inline]{\daphne, on the contrary, is designed to store a full history
%%%and, eventually, the whole (abstraction of the) transition system associated
%%%with the DCDS.}
%While such features can be disabled if not needed
%(thus yielding a simplified model), for the sake of completeness, we
%describe here the most general setting, where historical data are
%stored.

\begin{figure}[t]

\tikzstyle{data}=[thick,cylinder,draw,rotate=90,minimum width=8mm,minimum height=8mm]

\subfloat[][Enactment \label{fig:enactment}]{
  \scalebox{0.8}{
  \begin{tikzpicture}
    \node[data,fill=orange!10,densely dashed] (prev) at (0,0) {};
    \node at (0,0) {$\I$};
    \node[data,fill=red!20,ultra thick] (next) at (1.7,0) {};
    \node at (1.7,0) {$\I'$};
    \node[data,draw=none] at (0,-.6) {};
   
    \draw[-latex,ultra thick] (prev) edge (next);
    %\draw[-,ultra thick] (-.5,-.5) edge (.5,.5);
    %\draw[-,ultra thick] (prev.south east) edge (prev.north west);
  \end{tikzpicture}  
  }
}
\qquad
\subfloat[][Enactment with history recall \label{fig:history}]{
  \scalebox{0.8}{
  \begin{tikzpicture}
    \node[data,draw=white] at (0,-.6) {};
  
    \node[data,fill=orange!10] (t0) at (0,0) {};
    \node at (0,0) {$t_0$};
    \node[data,fill=red!20] (t1) at (1.2,0) {};
    \node at (1.2,0) {$t_1$};
    \node at (2.2,0) (dots) {$\ldots$};
    \node[data,fill=green!20] (tp) at (3.3,0) {};
    \node at (3.3,0) {$t_n$};
    \node[data,fill=blue!20, ultra thick] (tn) at (4.5,0) {};
    \node at (4.5,0) {$t_{n+1}$};
    
    \draw[-latex, very thick] (t0) edge (t1);
    \draw[-latex, very thick] (t1) edge (dots);
    \draw[-latex, very thick] (dots) edge (tp);
    \draw[-latex, ultra thick] (tp) edge (tn);
    %\draw[-,ultra thick] (-.5,-.5) edge (.5,.5);
    %\draw[-,ultra thick] (prev.south east) edge (prev.north west);
  \end{tikzpicture}  
  }
}
\qquad
\subfloat[][State space construction\label{fig:ss}]{
  \scalebox{0.8}{
  \begin{tikzpicture}
    \node[data,fill=orange!10] (s0) at (0,0) {};
    \node at (0,0) {$s_0$};
    \node[data,fill=red!20] (s1) at (1.7,.6) {};
    \node at (1.7,.6) {$s_1$};
    \node[data,fill=green!20] (s2) at (1.7,-.6) {};
    \node at (1.7,-.6) {$s_2$};
    \node[data,fill=blue!20] (s3) at (3.4,-.6) {};
    \node at (3.4,-.6) {$s_3$};
    \node (s4) at (3.4,.6) {\ldots}; 
    
    \draw[-latex, very thick] (s0) edge (s1);
    \draw[-latex, very thick] (s0) edge (s2);
    \draw[-latex, very thick] (s1) edge (s2);
    \draw[-latex, very thick] (s2) edge (s3);
    \draw[-latex, very thick] (s3) edge[bend right=20] (s2);
    \draw[-latex, very thick] (s1) edge (s4);
    \draw[-latex, very thick] (s3) edge (s4);

    %\draw[-,ultra thick] (-.5,-.5) edge (.5,.5);
    %\draw[-,ultra thick] (prev.south east) edge (prev.north west);
  \end{tikzpicture}  
  }
}
\caption{The three main usage modalities for \daphne, and sketch of the corresponding data structures stored within the DBMS}
\end{figure}
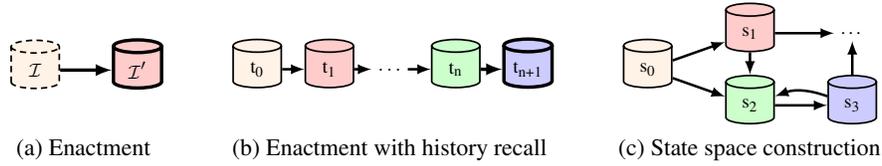

\subsection{Encoding a \dapsl in \daphne}
\label{sec:representation}

%%A DCDS specification consists of the definition of
%%a set of tables, accounting for the data layer, and a set of stored procedures,
%%together with auxiliary tables, implementing the DCDS process layer.
%%%%
%%Essentially, the DCDS is executed by running, through the
%%\fengine, the stored procedures on the input schema.
%%
%%The input specification is provided in terms of the native languages of
%%the chosen RDBMS for the definition of the data structure and the stored
%%procedures. Currently, it is responsibility of the modeler to design
%%an input specification that conforms to the \daphne format described
%%below. The development of a compiler which
%%takes care of this starting from a high-level input specification, is planned as future work.
%%%%

We now detail how \daphne encodes a \dapsl model
 \dapsl model $\sys = \tup{\dl,\pl}$ with data layer $\dl = \tup{\R, \E}$ and control layer $\pl = \tup{\F,\A,\rho}$ into a DBMS. Intuitively \daphne represents $\dl$ as a set of tables, and $\pl$ as a set of stored procedures working over those and auxiliary tables. 
%In \daphne, a $\dapl$ specification is represented through a set of tables, accounting for the data layer, and a set of stored procedures and additional auxiliary tables, realizing the process layer. 
%To execute a DCDS, one essentially needs to run (through the \fengine) the set of stored procedures on the data schema. 
Such data structures and stored procedures are defined in terms of the native language of the chosen DBMS. These can be either created manually, or automatically instrumented by \daphne itself once the user communicates to \daphne the content of $\sys$ using dedicated JAVA APIs. Specifically, we employ the jOOQ framework\footnote{\scriptsize \url{https://www.jooq.org/}} as the basis for the concrete input syntax of \dapsl models within \daphne. 
The interested reader may refer to Appendix~\ref{sec:interaction} to have a glimpse about how jOOQ and the DAPHNE APIs work.

Before entering into the encoding details, it is important to stress that \daphne provides three main usage modalities.
The first modality is \emph{enactment}. Here \daphne supports users in the process execution, storing the \emph{current DB}, and suitably updating it in response to the execution of actions.
The second modality is \emph{enactment with historical recall}. This is enactment where \daphne does not simply store the current state and evolves it, but also recalls historical information about the previous state configurations, i.e., the previous DBs, together with information about the applied actions (name, parameters, service call invocations and results, and timestamps). This provides full traceability about how the process execution evolved the initial state into the current one.
The last modality is \emph{state space construction for formal analysis}, where \daphne generates all possible ``relevant" possible executions of the system, abstracting away from timestamps and consequently folding the so-obtained traces into a \emph{relational transition system} \cite{Vard05,CDMP17}. Differently from the previous modalities, in this case \daphne does not simply account for a system run, but for the branching behaviour of $\sys$. %More details are given in Section~\ref{sec:architecture}.

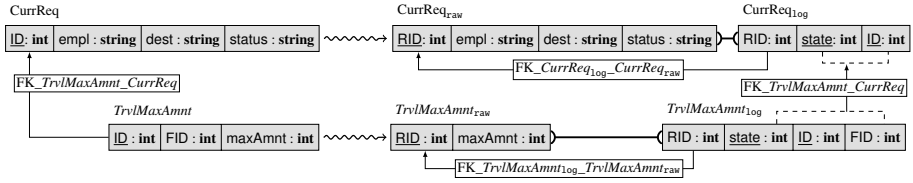
\begin{figure}[tb]
\centering
\resizebox{.99\hsize}{!}{
\begin{tikzpicture}[relation/.style={rectangle split, rectangle split parts=#1, rectangle split part align=base, draw, anchor=center, align=center, text height=3mm, text centered}]\label{fig:ex.tables.histroical.schema}%
%%%%%%%%%%%%%%%%%%%%%%%%%%%%%%%%%%
\hspace*{-0.3cm}
 \node (currreq_name) {$CurrReq$};
\node [relation=4, rectangle split horizontal, rectangle split part fill={lightgray!50}, below=0.6cm of currreq_name.west, anchor=west] (currreq)
{\underline{\attrname{ID}}: \typename{int}%
\nodepart{two} \attrname{empl} : \typename{string}
\nodepart{three}  \attrname{dest} : \typename{string}
\nodepart{four} \attrname{status} : \typename{string}};
%%%%%%%%%%%%%%%%%%%%%%%%%%%%%%%%%%

\node  [anchor=west,right=7.3cm of currreq_name.east](currreq_raw_name) {$CurrReq_\raw$};
\node [relation=4, rectangle split horizontal, rectangle split part fill={lightgray!50}, anchor=north west, below=0.6cm of currreq_raw_name.west, anchor=west] (currreq_raw)
{\underline{\attrname{RID}}: \typename{int}%
\nodepart{two} \attrname{empl} : \typename{string}
\nodepart{three}  \attrname{dest} : \typename{string}
\nodepart{four} \attrname{status} : \typename{string}};

%%%%%%%%%%%%%%%%%%%%%%%%%%%%%%%%%%

\node [anchor=west,right = 6cm of currreq_raw_name.east] (currreq_log_name) {$CurrReq_\slog$};
 \node [relation=3, rectangle split horizontal, rectangle split part fill={lightgray!50}, anchor=north west, below=0.6cm of currreq_log_name.west, anchor=west] (currreq_log)
{\attrname{RID}: \typename{int}%
\nodepart{two} \underline{\attrname{state}}: \typename{int}%
  \nodepart{three} \underline{\attrname{ID}}: \typename{int}};
%%%%%%%%%%%%%%%%%%%%%%%%%%%%%%%%%%
%%%%%%%%%%%%%%%%%%%%%%%%%%%%%%%%%%

\node [xshift=2.3cm,below=2.2cm of currreq_name.west, anchor=west] (trvlmaxamnt_name) {\relname{TrvlMaxAmnt}};
\node [relation=3, rectangle split horizontal, rectangle split part fill={lightgray!50}, anchor=north west, below=0.6cm of trvlmaxamnt_name.west, anchor=west] (trvlmaxamnt)
{\underline{\attrname{ID}} : \typename{int}%
\nodepart{two}  \attrname{FID} : \typename{int}
\nodepart{three}  \attrname{maxAmnt} : \typename{int}};
%%%%%%%%%%%%%%%%%%%%%%%%%%%%%%%%%%

\node [xshift=0cm,right=4.3cm of trvlmaxamnt_name.east, anchor=west] (trvlmaxamnt_raw_name) {$\relname{TrvlMaxAmnt}_\raw$};
\node [relation=2, rectangle split horizontal, rectangle split part fill={lightgray!50}, anchor=north west, below=0.6cm of trvlmaxamnt_raw_name.west, anchor=west] (trvlmaxamnt_raw)
{\underline{\attrname{RID}} : \typename{int}%
\nodepart{two}  \attrname{maxAmnt} : \typename{int}};
%%%%%%%%%%%%%%%%%%%%%%%%%%%%%%%%%%

\node [xshift=0cm,right=3.7cm of trvlmaxamnt_raw_name.east, anchor=west] (trvlmaxamnt_log_name) {$\relname{TrvlMaxAmnt}_\slog$};
\node [relation=4, rectangle split horizontal, rectangle split part fill={lightgray!50}, anchor=north west, below=0.6cm of trvlmaxamnt_log_name.west, anchor=west] (trvlmaxamnt_log)
{\attrname{RID} : \typename{int}%
\nodepart{two}  \underline{\attrname{state}} : \typename{int}%
\nodepart{three}  \underline{\attrname{ID}} : \typename{int}%
\nodepart{four}  \attrname{FID} : \typename{int}};
%%%%%%%%%%%%%%%%%%%%%%%%%%%%%%%%%%

%%%%%%%%%%%%%%%%%%%%%%%%%%%%%%%%%%%
%%% ARROWS
%%%%%%%%%%%%%%%%%%%%%%%%%%%%%%%%%%%
%%% foreign keys
\draw[-latex] (currreq_log.one south) -- ++ (0,-0.40) -| node[rectangle,draw=black,fill=white,xshift=40mm,inner sep=.5mm,minimum size=1mm]{\footnotesize{\textup{FK\_}$\relname{CurrReq}_\slog$\textup{\_}$\relname{CurrReq}_\raw$}} ($(currreq_raw.two south) + (-1.7,0)$);
\draw[-latex] (trvlmaxamnt_log.one south) -- ++ (0,-0.40) -| node[rectangle,draw=black,fill=white,xshift=32mm,inner sep=.5mm,minimum size=1mm]{\footnotesize{\textup{FK\_}$\relname{TrvlMaxAmnt}_\slog$\textup{\_}$\relname{TrvlMaxAmnt}_\raw$}} ($(trvlmaxamnt_raw.two south) + (-1.7,0)$);
\draw[-latex] ($(trvlmaxamnt_log.north)+(1.396,0.27)$) -- node[rectangle,draw=black,fill=white,xshift=-5mm,inner sep=.5mm,minimum size=1mm]{\footnotesize{\textup{FK\_}$\relname{TrvlMaxAmnt}$\textup{\_}$\relname{CurrReq}$}}  ($(currreq_log.south)+(0.5,-0.27)$);
\draw[-latex] (trvlmaxamnt.one west) -- ++ (-0.50,0) -| node[rectangle,draw=black,fill=white,xshift=15mm,yshift=12mm,inner sep=.5mm,minimum size=1mm]{\footnotesize{\textup{FK\_}$\relname{TrvlMaxAmnt}$\textup{\_}$\relname{CurrReq}$}} ($(currreq.one south) + (0,0)$);
												
%%% connectors
\draw[line width=1.2pt,)-(] (currreq_raw) -- (currreq_log);
\draw[line width=1.2pt,)-(] (trvlmaxamnt_raw) -- (trvlmaxamnt_log);
\draw[dashed] ($(currreq_log.two south)+(-0.15,0)$) -- ++ (0,-0.25) -| ($(currreq_log.three south) + (0.05,0)$);
\draw[dashed] ($(trvlmaxamnt_log.two north)+(0.4,0)$) -- ++ (0,0.28) -| ($(trvlmaxamnt_log.four north) + (0.2,0)$);
%%% misc
\draw [->,decorate, thick,
decoration={snake,amplitude=.4mm,segment length=2mm,post length=1mm}] ($(currreq.east)+(0.1,0)$) -- ++ (1.4,0) (currreq_raw.west);
\draw [->,decorate, thick,
decoration={snake,amplitude=.4mm,segment length=2mm,post length=1mm}] ($(trvlmaxamnt.east)+(0.1,0)$) -- ++ (1.4,0) (trvlmaxamnt_raw.west);
\end{tikzpicture}}
\caption{Relational schemas of $\relname{CurrReq}$ and $\relname{TrvlMaxAmnt}$,  and their historical representation in \daphne via two pairs of corresponding tables: $\relname{CurrReq}_\raw$ and $\relname{CurrReq}_\slog$,  $\relname{TrvlMaxAmnt}_\raw$ and $\relname{TrvlMaxAmnt}_\slog$.}
\label{fig:ex.tables.historical}
\end{figure}

%In the following, we assume that all primary keys,
%and thus foreign keys,
%are single-attributed. This is only done for presentation pursposes.
%
%, $\id$ to refer to its primary keys,
%and notation $\fid_i$ to refer its $i$-th foreign keys (assuming to have fixed an ordering over foreign keys).
%\footnote{Here
% denotes a 
%projection of $\relname{R}$ on the ordered set of attributes , while $\FK{\relname{S}}{\attrname{B}_1,\ldots,\attrname{B}_K}
%{\relname{R}}{\attrname{A}_1,\ldots,\attrname{A}_m}$ defines a foreign key relation
% between two projections.}) 

\medskip\noindent \textbf{Data layer}.
\daphne does not internally store the data layer as it is specified in $\dl$, but adopts a more sophisticated schema. This is done to have a unique homogeneous approach that supports the three usage modalities mentioned before. In fact, instructing the DBMS to directly store the schema expressed in $\dl$ would suffice only in the enactment case, but not to store historical data about previous states, nor the state space with its branching nature.
To accommodate all three usages at once, \daphne proceeds as follows. Each relation schema $\relname{R}$ of $\dl$ becomes relativized to a \emph{state identifier}, and decomposed into two interconnected relation schemas:
\begin{inparaenum}[\it (i)]
\item $\relname{R}_\raw$ (\emph{raw data storage}), an inflationary table that incrementally stores all the tuples that have been ever inserted in $\relname{R}$; 
\item $\relname{R}_\slog$ (\emph{state log}), which is responsible at once for maintaining the referential integrity of the data in a state, as well as for fully reconstructing the exact  
 content of $\relname{R}$ in a state.  
\end{inparaenum}
In details, $\relname{R}_\raw$ contains all the attributes $A$ of
$\relname{R}$ that are \emph{not} part of primary keys nor sources of a foreign key, plus an additional surrogate identifier $\rid$, so that $\pk{\relname{R}_\raw} = \tup{\rid}$. Each possible combination of values over $A$ is stored only once in $\relname{R}_\raw$ (i.e., $\proj{\relname{R}_\raw}{A}$ is a key), thus maximizing compactness.
At the same time, $\relname{R}_\slog$ contains the following attributes:
\begin{inparaenum}[\it (i)]
\item an attribute $\logstate$ representing the state identifier;
\item the primary key of (the original relation) $R$; 
\item a reference to $\relname{R}_\raw$, i.e., an attribute $\rid$ with 
 $\FK{\relname{R}_\slog}{\rid}{\relname{R}_\raw}{\rid}$;
\item all attributes of $\relname{R}$ that are sources of a foreign key in $\dl$.
\end{inparaenum}
To guarantee referential integrity, $\relname{R}_\slog$ must ensure that (primary) keys and foreign keys are now relativized to a state. This is essential, as the same tuple of $R$ may evolve across states, consequently requiring to historically store its different versions, and suitably keep track of which version refers to which state. Also foreign keys have to be understood within the same state: if a reference tuple changes from one state to the other, all the other tuples referencing it need to update their references accordingly. To realize this, we set $\pk{R_\slog} = \tup{\pk{R},\logstate}$. Similarly, for each foreign key $\FK{\relname{S}}{B}{\relname{R}}{A}$ originally associated to relations $\relname{R}$ and $\relname{S}$ in $\dl$, we insert in the DBMS the foreign key $\FK{\relname{S}_\slog}{B,\logstate}{\relname{R}_\slog}{A,\logstate}$ over their corresponding state log relations.

With this strategy, the ``referential" part of $\relname{R}$ is suitably relativized w.r.t.~a state, while at the same time all the other attributes are compactly stored in $\relname{R}_\raw$, and referenced possibly multiple times from $\relname{R}_\slog$. In addition, notice that, given a state identified by $\cname{s}$, the full extension of relation $\relname R$ in $\cname{s}$ can be fully reconstructed by
\begin{inparaenum}[\it(i)]
\item selecting the tuples of $\relname{R}_\slog$
where~$\logstate=\cname{s}$;
\item joining the obtained selection with $\relname{R}_\raw$ on
$\rid$;
\item finally projecting the result on the original attributes of $\relname{R}$.
\end{inparaenum}
In general, this technique shows how an arbitrary SQL query over $\dl$ can be directly reformulated as a state-relativized query over the corresponding \daphne schema.

\begin{figure}[t]
  \centering
\resizebox{.9\hsize}{!}{
\begin{tikzpicture}
\begin{scope}
  \matrix (currreq1_raw) [table,text width=4.58em]
          {
            RID  \& Empl \& Dest \& Status \\
            	--  \& -- \& -- \& -- \\
          };
   \node (name_currreq1_raw)  at ([yshift=2mm,xshift=-20mm]currreq1_raw.north) {\Large{$\relname{CurrReq}_\raw$}};

  \matrix (currreq1_log) [table,text width=2.7em,xshift=6.0cm]
          {
            state \& ID \& RID  \\
                --\& -- \& -- \\
          };
   \node (name_currreq1_log)  at ([yshift=2mm,xshift=-4mm]currreq1_log.north) {\Large{$\relname{CurrReq}_\slog$}};
   
      %%%%%%%%% Pending
   
   \matrix (pending1_raw) [table,text width=4.7em, yshift = -2.9cm,xshift=0.8cm]
          {
           RID    \& Empl \& Dest  \\
           3  \& Bob \& NY \\
           4 \& Kriss \& Paris \\
          };
   \node (name_pending1_raw)  at ([yshift=2mm,xshift=-13mm]pending1_raw.north) {\Large{$\relname{Pending}_\raw$}};

    \matrix (pending1_log) [table,text width=2.7em, right of = pending1_raw, xshift= 4.2cm, yshift = 0.018cm]
          {
            state \& ID \& RID  \\
           1 \& 1 \& 3\\
           1 \& 2 \& 4\\
          };
   \node (name_pending1_log)at ([yshift=2mm,xshift=-5mm]pending1_log.north) {\Large{$\relname{Pending}_\slog$}};
\end{scope}

\begin{scope}[xshift=14.0cm]

   %%%%%%%%% CurrReq
   
 \matrix (currreq2_raw) [table,text width=4.58em]
          {
            RID  \& Empl \& Dest \& Status \\
            	5  \& Kriss \& Paris \& submitd \\
          };
   \node (name_currreq2_raw)  at ([yshift=2mm,xshift=-20mm]currreq2_raw.north) {\Large{$\relname{CurrReq}_\raw$}};

  \matrix (currreq2_log) [table,text width=2.7em,xshift=6.0cm]
          {
            state \& ID \& RID  \\
                2\& 2 \& 5 \\
          };
   \node (name_currreq2_log)  at ([yshift=2mm,xshift=-4mm]currreq2_log.north) {\Large{$\relname{CurrReq}_\slog$}};
   
      %%%%%%%%% Pending
   
   \matrix (pending2_raw) [table,text width=4.7em, yshift = -2.6cm,xshift=0.8cm]
          {
           RID    \& Empl \& Dest  \\
           3  \& Bob \& NY \\
           4 \& Kriss \& Paris \\
          };
   \node (name_pending2_raw)  at ([yshift=2mm]pending2_raw.north) {\Large{$\relname{Pending}_\raw$}};

    \matrix (pending2_log) [table,text width=2.7em, right of = pending2_raw, xshift= 4.2cm, yshift = 0.018cm]
          {
            state \& ID \& RID  \\
           1 \& 1 \& 3\\
           1 \& 2 \& 4\\
           2\& 1 \&3\\
          };
   \node (name_pending2_log)at ([yshift=2mm,xshift=-5mm]pending2_log.north) {\Large{$\relname{Pending}_\slog$}};
\end{scope}

%%%%%% LINES %%%%%%
\draw[line width=1.5pt,)-(] ($(currreq1_raw)+(3.33,0)$) -- ($(currreq1_log)+(-1.6,0)$);
\draw[line width=1.5pt,)-(] ($(currreq2_raw)+(3.33,0)$) -- ($(currreq2_log)+(-1.6,0)$);
\draw[line width=1.5pt,)-(] ($(pending1_raw)+(2.55,0)$) -- ($(pending1_log)+(-1.60,0)$);
\draw[line width=1.5pt,)-(] ($(pending2_raw)+(2.55,0)$) -- ($(pending2_log)+(-1.60,0)$);

%%%%%%% STATES %%%%%%%%%%
\node[rectangle,draw=black,dashed,fill=yellow!15,inner sep=.5mm,minimum size=1mm] (s1) at (-3.4,-3.6){\large{\cname{state\quad 1}}};
\node[rectangle,draw=black,dashed,fill=yellow!15,inner sep=.5mm,minimum size=1mm] (s2) at (10.4,-3.6){\large{\cname{state\quad 2}}};

\node[state] (a1) [above of=name_currreq1_raw,draw=none,node distance=.7cm,yshift=-.7cm,xshift=-1.3cm] {};
\node[state] (a2) [below of=currreq1_log,draw=none,node distance=.2cm,xshift=1.7cm,yshift=-3.3cm] {};
\node[state] (b1) [above of=name_currreq2_raw,draw=none,node distance=.2cm,yshift=-.2cm,xshift=-1.3cm] {};
\node[state] (b2) [below of=currreq2_log,draw=none,node distance=.2cm,xshift=1.7cm,yshift=-3.3cm] {};

\node[rectangle,draw=black,fill=white!5,inner sep=.5mm,minimum size=1mm] (bidning) at (9.2,-1.2)
{\large{$
\,\actname{StartWorkflow}(\cname{2},\cname{Kriss},\cname{Paris})\,$}};
\node[state,draw=none,node distance=.2cm] (c1) at (5.8,-1.2){};
\node[state,draw=none,node distance=.2cm] (c2) at (13.2,-1.2){};

%\node[state] (c1) [below of=pending1_log,draw=none,node distance=.2cm,xshift=-0.6cm,yshift=-1.0cm] {};
%\node[state] (c2) [below of=pending2_raw,draw=none,node distance=.2cm,xshift=-1.4cm,yshift=-1.0cm] {};

\begin{pgfonlayer}{background}
\node [fill=blue!5,rounded corners,dashed,draw=black!50,fit=(a1) (a2)] {}; 
\node [fill=blue!5,rounded corners,dashed,draw=black!50,fit=(b1) (b2)] {}; 
\path[-latex,line width=4.5pt] (c1) edge node [left] {} (c2);

%\path[-latex,line width=1.5pt,every node/.style={font=\sffamily\small}]
%    (c1) edge[bend right] node [left] {} (c2);
\end{pgfonlayer}

\end{tikzpicture}}
\caption{Action application with two partial database snapshots mentioning $\relname{Pending}$ and $\relname{CurrReq}$. Here, $\actname{StartWorkflow}$ is applied in state \cname{1} with $\set{\pname{id}=\cname{2},\pname{empl}=\cname{Kriss},\pname{dest}=\cname{Paris}}$ as binding and, in turn, generates a new state \cname{2}.}
\label{fig:ex.tables.snapshot.1}
\end{figure}
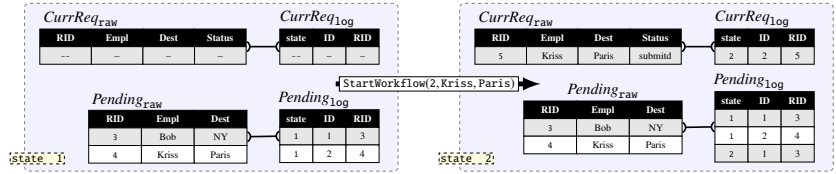

\begin{figure}[t]
  \centering
\resizebox{.99\hsize}{!}{
\begin{tikzpicture}
\begin{scope}
  \matrix (currreq1_raw) [table,text width=4.58em]
          {
            RID  \& Empl \& Dest \& Status \\
                5 \& Kriss \& Paris \& submitd \\
          };
   \node (name_currreq1_raw)  at ([yshift=2mm,xshift=-20mm]currreq1_raw.north) {\Large{$\relname{CurrReq}_\raw$}};

  \matrix (currreq1_log) [table,text width=2.7em,xshift=6.0cm]
          {
            state \& ID \& RID  \\
                2 \& 2 \& 5 \\
          };
   \node (name_currreq1_log)  at ([yshift=2mm,xshift=-4mm]currreq1_log.north) {\Large{$\relname{CurrReq}_\slog$}};
   
      %%%%%%%%% TrvlMaxAmnt
   
   \matrix (trvlmaxamnt1_raw) [table,text width=4.7em, yshift = -2.4cm,xshift=-1.65cm]
          {
           RID    \& maxAmnt  \\
            --  \& -- \\
          };
   \node (name_trvlmaxamnt1_raw)  at ([yshift=2mm]trvlmaxamnt1_raw.north) {\Large{$\relname{TrvlMaxAmnt}_\raw$}};

    \matrix (trvlmaxamnt1_log) [table,text width=2.7em, right of = trvlmaxamnt1_raw, xshift= 4.0cm, yshift = 0.018cm]
          {
            state \& ID \& RID \& FID \\
           -- \& -- \& --\& -- \\
          };
   \node (name_trvlmaxamnt1_log)at ([yshift=2mm]trvlmaxamnt1_log.north) {\Large{$\relname{TrvlMaxAmnt}_\slog$}};
\end{scope}

\begin{scope}[xshift=13.7cm]

   %%%%%%%%% CurrReq
   
 \matrix (currreq2_raw) [table,text width=4.58em]
          {
            RID  \& Empl \& Dest \& Status \\
                5 \& Kriss \& Paris \& submitd \\
                6 \& Kriss \& Paris \& acceptd \\
          };
   \node (name_currreq2_raw)  at ([yshift=2mm,xshift=-20mm]currreq2_raw.north) {\Large{$\relname{CurrReq}_\raw$}};

  \matrix (currreq2_log) [table,text width=2.7em,xshift=6.0cm]
          {
            state \& ID \& RID  \\
                2 \& 2 \& 5 \\
                3 \& 2 \& 6 \\
          };
   \node (name_currreq2_log)  at ([yshift=2mm,xshift=-4mm]currreq2_log.north) {\Large{$\relname{CurrReq}_\slog$}};
   
   %%%%%%%%% TrvlMaxAmnt
   
   \matrix (trvlmaxamnt2_raw) [table,text width=4.7em, yshift = -2.4cm, xshift = 1.37cm]
          {
           RID    \& maxAmnt  \\
           10  \& 900 \\
          };
   \node (name_trvlmaxamnt2_raw)  at ([yshift=2mm]trvlmaxamnt2_raw.north) {\Large{$\relname{TrvlMaxAmnt}_\raw$}};

    \matrix (trvlmaxamnt2_log) [table,text width=2.7em, right of = trvlmaxamnt2_raw, xshift= 4.0cm, yshift = 0.018cm]
          {
            state \& ID \& RID \& FID \\
           3 \& 3 \& 10 \& 2 \\
          };
   \node (name_trvlmaxamnt2_log)at ([yshift=2mm]trvlmaxamnt2_log.north) {\Large{$\relname{TrvlMaxAmnt}_\slog$}};
\end{scope}

%%%%%% LINES %%%%%%
\draw[line width=1.5pt,)-(] ($(currreq1_raw)+(3.33,0)$) -- ($(currreq1_log)+(-1.6,0)$);
\draw[line width=1.5pt,)-(] ($(currreq2_raw)+(3.33,0)$) -- ($(currreq2_log)+(-1.6,0)$);
\draw[line width=1.5pt,)-(] ($(trvlmaxamnt1_raw)+(1.71,0)$) -- ($(trvlmaxamnt1_log)+(-2.16,-0.02)$);
\draw[line width=1.5pt,)-(] ($(trvlmaxamnt2_raw)+(1.71,0)$) -- ($(trvlmaxamnt2_log)+(-2.16,-0.02)$);

%%%%%%% STATES %%%%%%%%%%
\node[rectangle,draw=black,dashed,fill=yellow!15,inner sep=.5mm,minimum size=1mm] (s1) at (-3.4,-3.3){\large{\cname{state\quad 2}}};
\node[rectangle,draw=black,dashed,fill=yellow!15,inner sep=.5mm,minimum size=1mm] (s2) at (10.3,-3.3){\large{\cname{state\quad 3}}};

\node[state] (a1) [above of=name_currreq1_raw,draw=none,node distance=.7cm,yshift=-.45cm,xshift=-1.3cm] {};
\node[state] (a2) [below of=currreq1_log,draw=none,node distance=.2cm,xshift=1.7cm,yshift=-2.9cm] {};
\node[state] (b1) [above of=name_currreq2_raw,draw=none,node distance=.2cm,yshift=-.2cm,xshift=-1.3cm] {};
\node[state] (b2) [below of=currreq2_log,draw=none,node distance=.2cm,xshift=2.4cm,yshift=-2.9cm] {};

%\node[state] (c1) [below of=trvlmaxamnt1_log,draw=none,node distance=.2cm,xshift=-0.8cm,yshift=-1.0cm] {};
%\node[state] (c2) [below of=trvlmaxamnt2_raw,draw=none,node distance=.2cm,xshift=-0.7cm,yshift=-1.0cm] {};
\node[state,draw=none,node distance=.2cm] (c1) at (5.6,-1.7){};
\node[state,draw=none,node distance=.2cm] (c2) at (13.2,-1.7){};
\node[rectangle,draw=black,fill=white!5,inner sep=.5mm,minimum size=1mm] (bidning) at (9.1,-1.7)
{\large{$\begin{array}{ll}
\actname{RvwRequest}(\cname{2},\cname{Kriss},\cname{Paris})\\
\overline{\scname{maxAmnt}(\cname{Krissl},\cname{Paris})=\cname{900}}
\end{array}$}};

\begin{pgfonlayer}{background}
\node [fill=blue!5,rounded corners,dashed,draw=black!50,fit=(a1) (a2)] {}; 
\node [fill=blue!5,rounded corners,dashed,draw=black!50,fit=(b1) (b2)] {}; 
%\path[-latex,line width=1.5pt,every node/.style={font=\sffamily\small}]
%    (c1) edge[bend right] node [left] {} (c2);
\path[-latex,line width=4.5pt] (c1) edge node [left] {} (c2);
\end{pgfonlayer}

\end{tikzpicture}}
\caption{Action application with two partial database snapshots mentioning $\relname{CurrReq}$ and $\relname{TrvlMaxAmnt}$. Here, $\actname{RvwRequest}$ is applied in state \cname{2} with $\set{\pname{id}=\cname{2},\pname{empl}=\cname{Kriss},\pname{dest}=\cname{Paris}}$ as binding and $\cname{900}$ resulting from the invocation to service $\scname{maxAmnt}$ that, in turn, generates a new state \cname{3}.}
\label{fig:ex.tables.snapshot.2}
\end{figure}
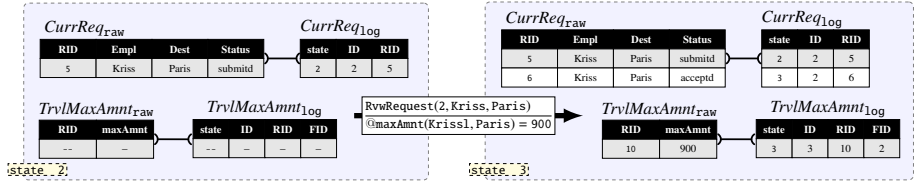

\begin{figure}[t]
  \centering
\resizebox{.99\hsize}{!}{
\begin{tikzpicture}
\begin{scope}
  \matrix (currreq1_raw) [table,text width=4.58em]
          {
            RID  \& Empl \& Dest \& Status \\
                6 \& Kriss \& Paris \& accepted \\
          };
   \node (name_currreq1_raw)  at ([yshift=2mm,xshift=-20mm]currreq1_raw.north) {\Large{$\relname{CurrReq}_\raw$}};

  \matrix (currreq1_log) [table,text width=2.7em,xshift=6.0cm]
          {
            state \& ID \& RID  \\
                3 \& 2 \& 6 \\
          };
   \node (name_currreq1_log)  at ([yshift=2mm,xshift=-4mm]currreq1_log.north) {\Large{$\relname{CurrReq}_\slog$}};
   
      %%%%%%%%% TrvlMaxAmnt
   
   \matrix (trvlmaxamnt1_raw) [table,text width=4.7em, yshift = -2.2cm,xshift=-1.65cm]
          {
           RID    \& maxAmnt  \\
            10  \& 900 \\
          };
   \node (name_trvlmaxamnt1_raw)  at ([yshift=2mm]trvlmaxamnt1_raw.north) {\Large{$\relname{TrvlMaxAmnt}_\raw$}};

    \matrix (trvlmaxamnt1_log) [table,text width=2.7em, right of = trvlmaxamnt1_raw, xshift= 4.0cm, yshift = 0.018cm]
          {
            state \& ID \& RID \& FID \\
           3 \& 3 \& 10 \& 2 \\
          };
   \node (name_trvlmaxamnt1_log)at ([yshift=2mm]trvlmaxamnt1_log.north) {\Large{$\relname{TrvlMaxAmnt}_\slog$}};

    %%%%%%%%% TrvlCost
   
   \matrix (trvlcost1_raw) [table,text width=4.7em, yshift = -4.3cm,xshift=-1.65cm]
          {
           RID    \& cost  \\
            --  \& -- \\
          };
   \node (name_trvlcost1_raw)  at ([yshift=2mm]trvlcost1_raw.north) {\Large{$\relname{TrvlCost}_\raw$}};

    \matrix (trvlcost1_log) [table,text width=2.7em, right of = trvlcost1_raw, xshift= 4.0cm, yshift = 0.018cm]
          {
            state \& ID \& RID \& FID \\
           -- \& -- \& -- \& -- \\
          };
   \node (name_trvlcost1_log)at ([yshift=2mm]trvlcost1_log.north) {\Large{$\relname{TrvlCost}_\slog$}};
\end{scope}

\begin{scope}[xshift=13.7cm]

   %%%%%%%%% CurrReq
   
 \matrix (currreq2_raw) [table,text width=4.58em]
          {
            RID  \& Empl \& Dest \& Status \\
                6 \& Kriss \& Paris \& acceptd \\
                7 \& Kriss \& Paris \& complete \\
          };
   \node (name_currreq2_raw)  at ([yshift=2mm,xshift=-20mm]currreq2_raw.north) {\Large{$\relname{CurrReq}_\raw$}};

  \matrix (currreq2_log) [table,text width=2.7em,xshift=6.0cm]
          {
            state \& ID \& RID  \\
                3 \& 2 \& 6 \\
                4 \& 2 \& 7 \\
          };
   \node (name_currreq2_log)  at ([yshift=2mm,xshift=-4mm]currreq2_log.north) {\Large{$\relname{CurrReq}_\slog$}};
   
   %%%%%%%%% TrvlMaxAmnt
   
   \matrix (trvlmaxamnt2_raw) [table,text width=4.7em, yshift = -2.6cm, xshift = 1.37cm]
          {
           RID    \& maxAmnt  \\
           10  \& 900 \\
          };
   \node (name_trvlmaxamnt2_raw)  at ([yshift=2mm]trvlmaxamnt2_raw.north) {\Large{$\relname{TrvlMaxAmnt}_\raw$}};

    \matrix (trvlmaxamnt2_log) [table,text width=2.7em, right of = trvlmaxamnt2_raw, xshift= 4.0cm, yshift = 0.018cm]
          {
            state \& ID \& RID \& FID \\
           3 \& 3 \& 10 \& 2 \\
           4 \& 3 \& 10 \& 2 \\
          };
   \node (name_trvlmaxamnt2_log)at ([yshift=2mm]trvlmaxamnt2_log.north) {\Large{$\relname{TrvlMaxAmnt}_\slog$}};

   %%%%%%%%% TrvlCost
   
   \matrix (trvlcost2_raw) [table,text width=4.7em, yshift = -4.89cm,xshift=1.37cm]
          {
           RID    \& cost  \\
            11 \& 700 \\
          };
   \node (name_trvlcost2_raw)  at ([yshift=2mm]trvlcost2_raw.north) {\Large{$\relname{TrvlCost}_\raw$}};

    \matrix (trvlcost2_log) [table,text width=2.7em, right of = trvlcost2_raw, xshift= 4.0cm, yshift = 0.018cm]
          {
            state \& ID \& RID \& FID \\
           4 \& 4 \& 11 \& 2 \\
          };
   \node (name_trvlcost2_log)at ([yshift=2mm]trvlcost2_log.north) {\Large{$\relname{TrvlCost}_\slog$}};
\end{scope}

%%%%%% LINES %%%%%%
\draw[line width=1.5pt,)-(] ($(currreq1_raw)+(3.33,0)$) -- ($(currreq1_log)+(-1.6,0)$);
\draw[line width=1.5pt,)-(] ($(currreq2_raw)+(3.33,0)$) -- ($(currreq2_log)+(-1.6,0)$);
\draw[line width=1.5pt,)-(] ($(trvlmaxamnt1_raw)+(1.71,0)$) -- ($(trvlmaxamnt1_log)+(-2.16,-0.02)$);
\draw[line width=1.5pt,)-(] ($(trvlmaxamnt2_raw)+(1.71,0)$) -- ($(trvlmaxamnt2_log)+(-2.16,-0.02)$);
\draw[line width=1.5pt,)-(] ($(trvlcost1_raw)+(1.71,0)$) -- ($(trvlcost1_log)+(-2.16,-0.02)$);
\draw[line width=1.5pt,)-(] ($(trvlcost2_raw)+(1.71,0)$) -- ($(trvlcost2_log)+(-2.16,-0.02)$);

%%%%%%% STATES %%%%%%%%%%
\node[rectangle,draw=black,dashed,fill=yellow!15,inner sep=.5mm,minimum size=1mm] (s1) at (-3.4,-5.45){\large{\cname{state\quad 3}}};
\node[rectangle,draw=black,dashed,fill=yellow!15,inner sep=.5mm,minimum size=1mm] (s2) at (10.3,-5.45){\large{\cname{state\quad 4}}};

\node[rectangle,draw=black,fill=white!5,inner sep=.5mm,minimum size=1mm] (bidning) at (9.1,-2.6)
{\large{$\begin{array}{ll}
\actname{FillReimb}(\cname{2},\cname{Kriss},\cname{Paris})\\
\overline{\scname{cost}(\cname{Krissl},\cname{Paris})=\cname{700}}
\end{array}$}};

\node[state] (a1) [above of=name_currreq1_raw,draw=none,node distance=.7cm,yshift=-.45cm,xshift=-1.3cm] {};
\node[state] (a2) [below of=currreq1_log,draw=none,node distance=.2cm,xshift=1.7cm,yshift=-5.0cm] {};
\node[state] (b1) [above of=name_currreq2_raw,draw=none,node distance=.2cm,yshift=-.2cm,xshift=-1.3cm] {};
\node[state] (b2) [below of=currreq2_log,draw=none,node distance=.2cm,xshift=2.4cm,yshift=-5.0cm] {};

%\node[state] (c1) [below of=trvlmaxamnt1_log,draw=none,node distance=.2cm,xshift=-0.8cm,yshift=-1.0cm] {};
%\node[state] (c2) [below of=trvlmaxamnt2_raw,draw=none,node distance=.2cm,xshift=-0.7cm,yshift=-1.0cm] {};
\node[state,draw=none,node distance=.2cm] (c1) at (5.6,-2.6){};
\node[state,draw=none,node distance=.2cm] (c2) at (13.1,-2.6){};
\node[rectangle,draw=black,fill=white!5,inner sep=.5mm,minimum size=1mm] (bidning) at (9.1,-2.6)
{\large{$\begin{array}{ll}
\actname{FillReimb}(\cname{2},\cname{Kriss},\cname{Paris})\\
\overline{\scname{cost}(\cname{Krissl},\cname{Paris})=\cname{700}}
\end{array}$}};

\begin{pgfonlayer}{background}
\node [fill=blue!5,rounded corners,dashed,draw=black!50,fit=(a1) (a2)] {}; 
\node [fill=blue!5,rounded corners,dashed,draw=black!50,fit=(b1) (b2)] {}; 
%\path[-latex,line width=1.5pt,every node/.style={font=\sffamily\small}]
%    (c1) edge[bend right] node [left] {} (c2);
\path[-latex,line width=4.5pt] (c1) edge node [left] {} (c2);
\end{pgfonlayer}

\end{tikzpicture}}
\caption{Action application with three partial DB snapshots mentioning  $\relname{CurrReq}$, $\relname{TrvlMaxAmnt}$ and $\relname{TrvlCost}$. Here, $\actname{FillReimb}$ is applied in state \cname{3} with $\set{\pname{id}=\cname{2},\pname{empl}=\cname{Kriss},\pname{dest}=\cname{Paris}}$ as binding and $\cname{700}$ resulting from the invocation of service $\scname{cost}$ that, in turn, generates a new state \cname{4}.}
\label{fig:ex.tables.snapshot.3}
\end{figure}
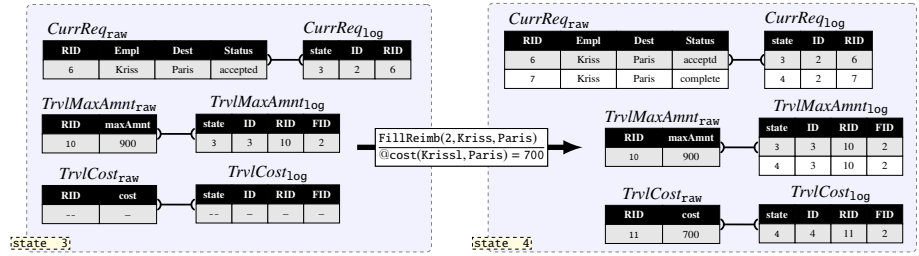

%%
%\todo[inline]{Fix the example according to Fig.4}
\begin{example}
Consider relation schemas $\relname{CurrReq}$ and $\relname{TrvlMaxAmnt}$  in Figure~\ref{fig:ex-bpm}. Figure~\ref{fig:ex.tables.historical} shows the representation of these relations in \daphne, suitably pairing $\relname{CurrReq}_\raw$ with
$\relname{CurrReq}_\slog$, and $\relname{TrvlMaxAmnt}_\raw$ with
$\relname{TrvlMaxAmnt}_\slog$. %Notice the presence of foreign keys:
Each state log table directly references a corresponding raw data storage table (e.g.,
$\FK{\relname{CurrReq}_\slog}{\attrname{RID}}{\relname{CurrReq}_\raw}{\attrname{RID}}$), 
and $\relname{TrvlMaxAmnt}$'s state log table, due to the FK in the original DAP, will reference a suitable key of $\relname{CurrReq}_\slog$ 
(i.e., $\FK{TrvlMaxAmnt_\slog}{\attrname{state},\attrname{FID}}{CurrReq_\slog}{\attrname{state},\attrname{ID}}$). 
Figures~\ref{fig:ex.tables.snapshot.1}, \ref{fig:ex.tables.snapshot.2} and \ref{fig:ex.tables.snapshot.3} show the evolution of the DBMS in response to the application of three ground actions, with full history recall. 
%concrete snapshots of $\relname{CurrReq}$ and $\relname{TrvlMaxAmnt}$ demonstrating historical data relevant to states $\cname{2}$ and $\cname{3}$. 
%
%Figure~\ref{fig:ex.tables.snapshot.2} depicts the state of $\relname{CurrReq}$,
%$\relname{TrvlCost}$, and $\relname{TrvlMaxAmnt}$, after the system has evolved
%through $4$ states (starting from $1$). Notice the presence of foreign keys:
%each state log table directly references a corresponding raw data storage table (e.g.,
%$\FK{\relname{CurrReq}_\slog}{\attrname{RID}}{\relname{CurrReq}_\raw}{\attrname{RID}}$), and other state log tables whenever an original foreign key existed in the DAP (e.g.,
%$\FK{TrvlMaxAmnt_\slog}{\attrname{state},\attrname{FID}}{CurrReq_\slog}{\attrname{state},\attrname{ID}}$).
\qedtriangle
\end{example}

We now discuss updates over $\relname{R}$.  As already pointed out,
$\relname R_\raw$ stores any tuple that occurs in some
state, that is, tuples are never deleted from $\relname R_\raw$.  Deletions are
simply obtained by \emph{not} referencing the deleted tuple in the new state.
%% For instance, in Figure~\ref{fig:ex.tables.snapshot}, it can be seen that
%% the first tuple of $\relname{CurrReq_\raw}$ (properly extended with its
%% $\id$, through $\rid$) has been deleted from $\relname{CurrReq}$ in state
%% $2$. It is indeed present in state $1$ (see first tuple of
%% $\relname{CurrReq}_\slog$) but not in state $2$ (see third tuple of
%% $\relname{CurrReq}_\slog$).
For instance, in Figure~\ref{fig:ex.tables.snapshot.1}, it can be seen that the
first tuple of $\relname{Pending_\raw}$ (properly extended with its $\id$,
through $\rid$) has been deleted from $\relname{Pending}$ in state $\cname{2}$:
while being present in state $\cname{1}$ (cf.~first tuple of
$\relname{Pending}_\slog$), the tuple is not anymore in state $\cname{2}$ (cf.~third
tuple of $\relname{Pending}_\slog$).

%For instance, in Figure~\ref{fig:ex.tables.snapshot}, it can be seen that the
%first tuple of $\relname{CurrReq_\raw}$ (properly extended with its $\id$,
%through $\rid$) has been deleted from $\relname{CurrReq}$ in state $2$: indeed,
%while being present in state $1$ (see first tuple of
%$\relname{CurrReq}_\slog$), the tuple is not anymore in state $2$ (see third
%tuple of $\relname{CurrReq}_\slog$).

%% As to additions, we proceed as follows.  Before inserting a new tuple, we
%% check whether it is already present in $\relname{R}_\raw$.  If so, we update
%% only $\relname{R}_\slog$ with a tuple referring, through $\rid$, that in
%% $\relname{R}_\raw$.  The value of attribute $\logstate$ depends on the state
%% that is being generated; the values of $\id$ and all $\fid_i$'s can be
%% obtained from any other tuple in $\relname{R}_\slog$ with same $\rid$.  If
%% the tuple is not present in $\relname{R}_\raw$, instead, it is added,
%% together with a fresh $\rid$.  Notice that its $\id$ and $\fid_i$'s are
%% provided as input, thus they are simply added, together with the value of
%% $\logstate$, when $\relname{R}_\slog$ is updated.
%%%%
%%In the actual implementation, $\relname{R}_\slog$ features also a $\hash$
%% attribute, with the value of a hash function computed based on the other
%% attributes (excluding $\rid$). This allows for checking quickly whether a
%% tuple is already present.

As for additions, we proceed as follows.  Before inserting a new tuple, we check
whether it is already present in $\relname{R}_\raw$.  If so, we update only
$\relname{R}_\slog$ by copying the $\relname{R}_\slog$ tuple referencing the
corresponding $\rid$ in $\relname{R}_\raw$.  In the copied tuple, the value of
attribute $\logstate$ is going to be the one of the newly generated state,
while the values of $\id$ and all foreign key attributes remain unchanged.  If the tuple
is not present in $\relname{R}_\raw$, it is also added to $\relname{R}_\raw$
together with a fresh $\rid$.  Notice that in that case its $\id$ and
FK attributes are provided as input, and thus they are simply added, together
with the value of $\logstate$, to $\relname{R}_\slog$.
In the actual implementation, $\relname{R}_\raw$ features also a $\hash$
attribute, with the value of a hash function computed based on original
$\relname{R}$ attributes (extracted from both $\relname{R}_\raw$ and
$\relname{R}_\slog$). This speeds up the search for identical tuples in $\relname{R}_\raw$.
%%}

%% Note that identifiers from $\key{R_\raw}$ uniquely define distinct tuples
%% per relation.  This allows us to perform updates in the presence of direct
%% or transitive dependencies between relations without disrupting temporal
%% integrity of the involved data.  For example, whenever we want to update
%% only $\relname{S}$ by changing some values in a row with a primary key $\id$
%% such that $\relname{R}[\fid]\subseteq \relname{S}[\id]$, our approach will
%% generate a new version of this row together with a new $\rid$ and insert a
%% corresponding new state tuple into $\relname{S}_\slog$.

%For unchanged relations $\relname S$, it is enough to update the corresponding
% relation $\relname S_\slog$ by copying previous state entries and updating
% their state identifiers' value to the actual one.  Notice that, if foreign
% keys are present, the pair $\logstate$-$\fid_i$ will reference the most
% recent versions of the previously referenced tuples.  As an example, consider
% again Figure~\ref{fig:ex.tables.snapshot}, where while \cname{Kriss} is
% changing his request status when moving from state \cname{2} to state
% \cname{3} (third and fourth lines of $\relname CurrReq_\slog$), the first
% tuple in $\relname{TrvlCost}$ references, through $\slog$ and $\fid$, the
% latest version of the tuple, i.e., the fourth one in
% $\relname{CurrReq}_\slog$.

Finally, we consider the case of relation schemas whose content is not changed when updating a state to a new state.  Assume
that relation schema $\relname{S}$ stays unaltered. After updating $\relname{R}$, it is enough
to update $\relname{S}_\slog$ by copying previous state entries and updating
the value of their state id to the actual one.  If a FK, whose left-hand side is $\proj{\relname{S}}{B}$, belongs to $\dl$ , the pair $\tup{B,\logstate}$ will reference the most recent versions of the previously referenced tuples.  Consider, e.g., 
Figure~\ref{fig:ex.tables.snapshot.3}. While in his current request \cname{Kriss} is changing the request status when moving from state \cname{3} to state \cname{4},
 the  maximum traveling budget assigned to this request (a tuple in 
 $\relname{TrvlMaxAmnt}$) should reference the latest version of the corresponding tuple in $\relname{CurrReq}$. Indeed, in state \cname{4}, a new tuple in $\relname{TrvlMaxAmnt}_\slog$ is referencing a new tuple in $\relname{CurrReq}_\slog$ that, in turn, corresponds to the one with the updated request status.

\medskip\noindent
\textbf{Control layer.} Each action \action of $\pl$, together with its dedicated CA rule, is encoded by \daphne into three stored procedures. The encoding is quite direct, thanks to the fact that both action conditions and action effect specifications are specified using SQL. 

The first stored procedure, \caeval, evaluates the CA rule of  \action in state
  \state over the respective DB (obtained by inspecting the state log relations whose state column matches with \state), and stores the all returned 
  parameter assignments for \action in a dedicated table \actparams. All parameter
  assignments in \actparams are initially unmarked, meaning that they are
  available for the user to choose.
  The second stored procedure, \effeval, executes queries corresponding to all the effects of \action
  over the DB of state \state, possibly using action parameters
  from \actparams extracted via a binding identifier \binding.
   Query results provide values to instantiate service
  calls, as well as those facts that must be deleted from or added to the current DB.  %Intermediate results are stored in auxiliary tables. 
The third stored procedure, {\effexec}, transactionally performs the actual delete and insert operations
  for a given state \state and a binding identifier \binding, using the results
  of service calls. % The domain constraints
%%  for the relations that are meant
%  to be updated are hard-coded in this stored procedure.

%
We describe now in detail the \daphne action execution cycle in a given state \state.
\begin{inparaenum}[\itshape (1)]
\item\label{cycle:start} The cycle starts with the user choosing one of the
  available actions presented by the \fengine. The available actions are
  acquired by calling \caeval, for each action \action in $\pl$.
\item\label{cycle:choice}
        If any unmarked parameter is present in \actparams,
        the user is asked to choose one of those (by selecting a binding identifier~\binding); once chosen, the
        parameter is marked, and the \fengine proceeds to the evaluation of \action by
        calling \effeval.
        If there are no such parameters, the user is asked to choose another available action, and
        the present step is repeated.
\item If \effeval involves service calls, these are passed to the 
        the \smanager component, which fetches the corresponding results.
\item\label{cycle:effect}
        \effexec is executed.  If all constraints in $\dl$ are satisfied, the
          transaction is committed and a new iteration starts from step~\ref{cycle:start};
        otherwise, the
          transaction is aborted and the execution history is kept unaltered, and
          the execution continues from step~\ref{cycle:choice}.
\end{inparaenum}

\subsection{Realization of the Three Usage Modalities}
\label{sec:modalities}
Let us now discuss how the three usage modalities are realized in \daphne. 
%With this strategy in place, the three usage modalities of \daphne are then  realized as follows. 
The simple enactment modality is realized by only recalling the current information about log relations. 
Enactment with history recall is instead handled as follows. First, the generation of a new state always comes with an additional update over an accessory 1-tuple relation schema indicating the timestamp of the actual update operation. The fact that timestamps always increase along the execution guarantees that each new state is genuinely different from previously encountered ones. Finally, an additional binary state transition table is employed, so as to keep track of the resulting total order over state identifiers. By considering our running example, in state $\cname{4}$ shown in Figure~\ref{fig:ex.tables.snapshot.3}, the content of the transition table would consist of the three pairs $\tup{\cname{1},\cname{2}}$, $\tup{\cname{2},\cname{3}}$, and $\tup{\cname{3},\cname{4}}$. 

As for state space construction, some preliminary observations are needed. Due to the presence of external services that may inject fresh input data, there are in general infinitely many different executions of the process, possibly visiting infinitely many different DBs (differing in at least one tuple). In other words, the resulting relational transition systems has infinitely many different states. However, thanks to the correspondence between \dapsl and DCDSs, we can realize in \daphne the abstraction techniques that have been introduced in \cite{BCDDM13,CDMP17} to attack the verification of such infinite-state transition systems. The main idea behind such abstraction techniques is the following. When carrying out verification, it is not important to observe all possible DBs that can be produced by executing the available actions with all possible service call results, but it suffices to only consider \emph{meaningful} combination of values, representing all possible ways to relate tuples with other tuples in the DB, in terms of (in)equality of their different components. This is done by carefully selecting the representative values. In \cite{BCDDM13,CDMP17}, it has been shown that this  technique produces a \emph{faithful} representation of the original relational transition system, and that this representation is also \emph{finite} if the original system is \emph{state bounded}, that is, has a pre-defined (possibly unknown) bound on the number of tuples that can be stored therein.\footnote{Even in the presence of this bound, infinitely many different DBs can be encountered, by changing the values stored therein.}  Constructing such a faithful representation is therefore the key towards analysis of formal properties such as reachability, liveness, and deadlock freedom, as well as explicit temporal model checking (in the style of \cite{CDMP17}).

State space construction is smoothly handled in \daphne as follows. When executed in this mode, \daphne replaces the service call manager with a mock-up manager that, whenever a service call is invoked, returns all and only \emph{meaningful} results, in the technical sense described above. E.g., if the current DB only contains string $\cval{a}$, invoking a service call that returns a string may only give two interesting results: $\cval{a}$ itself, or a string different than $\cval{a}$. To cover the latter case, the mock-up manager picks a representative value, say $\cval{b}$ in this example, implementing the representative selection strategy defined in \cite{BCDDM13,CDMP17}.
With this mock-up manager in place, \daphne constructs the state space by executing the following iteration. A state $\state$ is picked (at the beginning, only the initial state exists and can be picked). For each enabled ground action in $\state$, and for all \emph{relevant} possible results returned by the mock-up manager, the DB instance corresponding to the update is generated. If such a DB instance has been already encountered (i.e., is associated to an already existing state), then the corresponding state id $\state'$ is fetched. If this is not the case, a new id $\state'$ is created, inserting its content into the DBMS. Recall that $\state'$ is \emph{not} a timestamp, but just a symbolic, unique state id. The state transition table is then updated, by inserting the tuple $\tup{\state,\state'}$, which indeed witnesses that $\state'$ is one of the successors of $\state$.
The cycle is then repeated until all states and all enabled ground actions therein are processed. Notice that, differently from the enactment with history recall modality, in this case the state transition table is graph-structured, and in fact reconstructs the abstract representation of the relational transition system capturing the execution semantics of $\sys$. 

Figure~\ref{fig:ss} graphically depicts the state space constructed by \daphne on the travel reimbursement process whose initial DB only contains a single pending request. More detailed examples on the state space construction for the travel reimbursement process can be found in Appendix~\ref{sec:ss}.   

\begin{figure*}[t] 
\centering
\includegraphics[width=.9\textwidth]{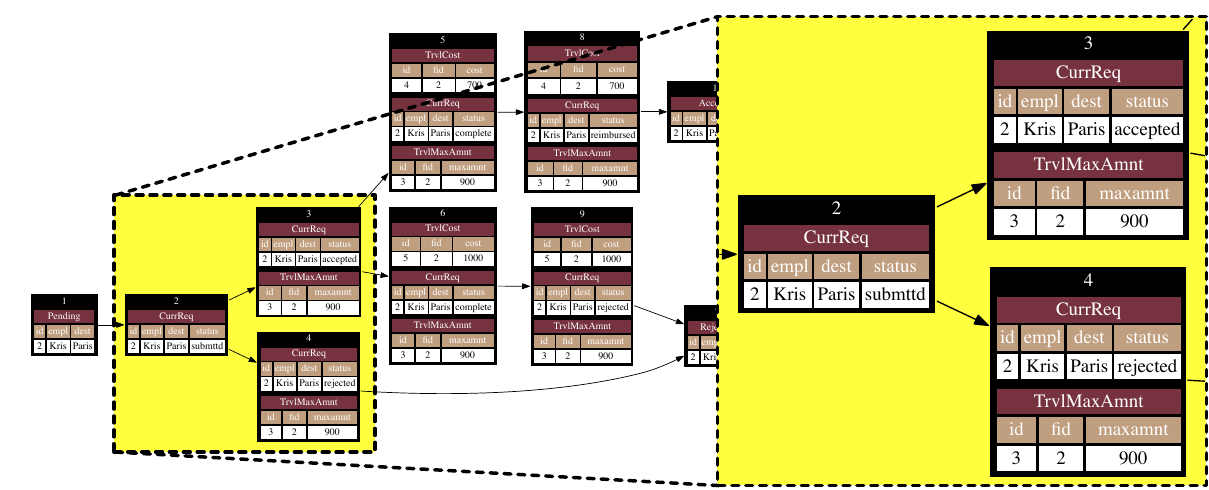}  
\caption{Example of state space constructed by \daphne}
\label{fig:ss}
\end{figure*}

%% file: discussion-and-related-work.tex
\section{Discussion and Related Work}
\label{sec:related-work}

%It has been widely acknowledged that conventional business
%process modeling methodologies fall short when
%considering complex data objects as tantamount modeling elements.
%To answer the increasing demand of integrated models holistically tackling the
%dynamics of a complex domain and the manipulation of
%data, a plethora of approaches have been proposed in both academia and industry.

Our approach directly relates to the family of data- and
artifact-centric approaches~\cite{CaDM13} that inspired the creation of various
modeling languages~\cite{Hull08,MeSW11} and execution frameworks such as:
\begin{inparaenum}[\itshape (i)]
\item the declarative rule-based \emph{Guard-Stage-Milestone} (GSM) language
  \cite{DaHV11} and its BizArtifact
  (\url{https://sourceforge.net/projects/bizartifact/}) execution platform;
\item the OMG \emph{CMMN} standard for case handling
  (\url{https://www.omg.org/spec/CMMN/});
\item the object-aware business process management approach implemented by
  \emph{PHILharmonic Flows}~\cite{KuWR11};
\item the extension of GSM called EZ-Flow \cite{XSYY11}, with SeGA as
  \cite{SuSY16} an execution platform;
\item the declarative data-centric process language \textsc{Reseda} based on
  term rewriting systems~\cite{Reseda18}.
\end{inparaenum}
As opposed to the more traditional activity-centric paradigms, these approaches emphasize the evolution of data objects through different states, but
often miss a clear representation of the control-flow dimension.
For example, GSM provides means for specifying business artifact lifecycles in
a declarative rule-based manner, and heavily relies on queries (ECA rules) over
the data to implicitly define the allowed execution flows.  
Similarly, our \dapsl language treats data as a ``first-class
citizen'' and allows one to represent data-aware business processes from the
perspective of database engineers.  
%Moreover, in our case data are also treated as the central modeling construct which alone drives the business process execution.
Other examples in \emph{(ii)}--\emph{(iv)} are rooted in similar abstractions to those of GSM, extending it towards more sophisticated interaction mechanisms
between artifacts and their lifecycle components.  

Our \dapsl language departs from these abstractions and provides a very pristine, programming-oriented solution that only retains the notions of data, actions, and CA rules. In this respect, the closes approach to ours among the aforementioned ones is \textsc{Reseda}. Similarly to \dapsl, \textsc{Reseda} in general allows one to specify reactive
rules and behavioral constraints defining the progression of a process in terms
of data rewrites and data inputs from outside the process context. 
\textsc{Reseda} manipulates only semi-structured data (such as XML or JSON)
that have to be specified directly in the tool. \dapsl focuses instead on purely relational data, and lends itself to be used on top of already specified relational DBs. 
%Instead, our framework
%% contrasts it by supporting
%supports relational data and, thanks to its low-level flavor, allows one to
%specify CA rules together with corresponding actions directly on top of an
%arbitrary database by using essentially only SQL. 
% The key idea is to associate either input events or reactive computation
% events to the individual elements of semi-structured data and declare
% reactive behaviour as explicit reaction rules and constraints between these
% events.  The data, along with the set of constraints, thereby at the same
% time constitutes the specification of the data, its behaviour and the
% run-time execution component.

% Nevertheless, many artifact-centric approaches do not come with
% well-established abstractions to capture the control-flow dimension, which
% tends in fact to stay implicit .

Differently from all such approaches, state-of-the-art business process management systems, along with an explicit representation of the process control flow, often provide sophisticated,
ad-hoc conceptual abstractions to manipulate business data. Notable examples of such systems are the Bizagi BPM suite (\url{bizagi.com}),
Bonita BPM (\url{bonitasoft.com}), Camunda (\url{camunda.com}), Activiti
(\url{activiti.org}), and YAWL (\url{yawlfoundation.org}).
%\footnote{\url{bizagi.com}, \url{bonitasoft.com}, \url{camunda.com},
%\url{activiti.org}, \url{yawlfoundation.org}}
%Even though such tools slightly vary in their modeling choices (e.g., in the
%graphical modeling language) and advanced functionalities, they share two
%common features: business processes definition languages are based on BPMN, and
%business data accessibility is managed via typed variables assigned to the
%workflow.  While many tools provide sophisticated data structures (e.g., tables
%for historical and runtime data) and storage options (e.g., in-memory or
%shared), the way the data are modified is often hidden inside the task
%logic. 
While they all provide an explicit representation of the process control flow using similar, well-accepted abstractions, they typically consider the task logic and its interaction with persistent data as a sort of  ``procedural attachment'', i.e., a piece of code whose functioning is not conceptually revealed \cite{CaDM13}. This is also apparent in standard languages such as BPMN, which consider the task and the decision logics as black boxes. The shortcomings of this assumption  have been extensively discussed in the literature \cite{Rich10,Dum11,Reic12,CaDM13}.
\dapsl, due to its pristine, data-centric flavor, can be used to complement such approaches with a declarative, explicitly exposed specification of the task and decision logic, using the well-accepted SQL language as main modeling metaphor.

%% file: conclusion.tex
\section{Conclusions}
\label{sec:conclusions}
We have introduced a declarative, purely relational framework for data- aware processes, in which SQL is used as the core data inspection and update language. We have reported on the implementation of this framework in DAPHNE, a system prototype grounded in standard relational technology and Java that at once accounts for modeling, enactment, and state space construction for verification.
As for modeling, we intend to interface DAPHNE with different concrete end user- oriented languages for the integrated modeling of processes and data, incorporating at once artifact- and activity-centric approaches. Since our approach is having a minimalistic, SQL-centric flavor, it would be also interesting to empirically validate it and, in particular, to study its usability among database experts who need to model processes. Daphne could in fact allow them to enter into process modeling by using a metaphor that is closer to their expertise. As for formal analysis, we plan to augment the state space construction with native verification capabilities to handle basic properties such as reachability and liveness, and even more sophisticated temporal logic model checking. At the moment, the development of verification tools for data- aware processes is at its infancy, with a few existing tools \cite{DDG16,LiDV17}. Finally, given that DAPHNE can generate a log including all performed actions and data changes, we aim at investigating its possible applications to process mining, where emerging trends are moving towards multi-perspective analyses that consider not just the sequence of events but also the corresponding data.

%%% Local Variables:
%%% mode: latex
%%% TeX-master: "main"
%%% End:

%% file: appendix.tex
\section{Interacting with \daphne}

\label{sec:interaction}
\daphne comes with dedicated APIs to acquire a \dapm, injects it in the underlying DBMS, and interact with the execution engine, on the one hand inspect the state of the process and the current data, and on the other to enact its progression. In order to avoid redundant technicalities and still give a flavor of the way \daphne 
works, we show how to specify and enact {\dapm}s on top of it for the case when the history recall modality is chosen.

As a concrete specification language for \dapm, we are developing an extension of the standard SQL, mixed with few 
syntactic additions allowing to compactly define \dapm rules and actions. 
Concretely, such a language is directly embedded into Java using the jOOQ framework\footnote{\scriptsize \url{https://www.jooq.org/}}, 
which allows to create and manipulate relational tables, constraints, and SQL queries as Java objects.
Building on top of the JOOQ APIs, we have realized an additional API layer that, given a jOOQ object describing a \dapm component (such as a relation, rule, or action), automatically transforms it into a corresponding SQL or PL/pgSQL code snippet following the strategy defined in Section~\ref{sec:representation}. 
E.g., one could specify
the $\relname{CurrReq}$ schema and its primary key using the following code:
\begin{Java}
DSLContext create = DSL.using(SQLDialect.POSTGRES_9_4);
Field<Integer> id = DSL.field(DSL.name("CurrReq", "id"),
	SQLDataType.INTEGER.nullable(false).identity(true));
Field<String> empl = DSL.field(DSL.name("CurrReq", "empl"),
	SQLDataType.VARCHAR.length(40).nullable(false));
Field<String> dest = DSL.field(DSL.name("CurrReq", "dest"),
	SQLDataType.VARCHAR.length(40).nullable(false));
Field<String> status = DSL.field(DSL.name("CurrReq", "status"),
	SQLDataType.VARCHAR.length(40).nullable(false));
Schema currReq = new Schema("CurrReq", id, empl, dest, status);
PrimaryKey pkCurrReq = new PrimaryKey("pk_CurrReq", currReq,
	Arrays.asList(id));
currReq.addConstraint(pkCurrReq);
\end{Java}
To generate the SQL code for that relation, which should account for creating the state log and raw data storage relations for $\relname{CurrReq}$ (and related constraints), one can then either rely on conventional JDBC methods, or directly invoke the {\daphne} API:
\begin{Java}
String createCurrReq_script = currReq.getSQLTranslation(create);
\end{Java}
The following snippet shows how to create a CA rule for action $\actname{RvwRequest}$:
\begin{Java}
Select query = create.select(id,empl, dest)
	.from(currReq.generateTable(), trvlCost.generateTable())
	.where(status.eq("submitted"));
Action rvwRequest = new Action("RvwRequest");
CARule rvwRequestCARule = new CARule(query, rvwRequest);
\end{Java}
To generate a script that will instrument the necessary tables and stored procedures in the underlying \dapm, as stipulated in Section~\ref{sec:representation}, the user just needs to call another dedicated \daphne API:
\begin{Java}
String script = rvwRequestCARule.getSQLTranslation(create);
\end{Java}
%%

%==
Let us now provide a flavour of the runtime API of \daphne, employed during the enactment of a system run.  First, we need to obtain an instance of process engine, providing its corresponding database connection details:
\begin{Java}
Connection con = DBTools.getConnection(...);
StatefulEngine dcds = StatefulEngine.getEngingeInstance(con);
\end{Java}
The engine provides a plethora of state inspection and update functionalities. We show how to create an action provider object that provides metadata (such as action names and their params) about all actions that are enabled in the current state.
\begin{Java}
State currState = dcds.getCurrentState();
ActionProvider provider = dcds.getActionProvider(currState);
List<Attributes> md = provider.getAvailableActionsMetaData();
\end{Java}
Using one of the so-obtained metadata, we create an action and a binding provider, in turn used to obtain all the current legal parameter instantiations for such an action.
\begin{Java}
Attributes att = md.get(0);
ImmutablePair<Action, BindingProvider> actionElements
	= provider.getAction(att);
\end{Java}
Among the available bindings, we pick one, getting back a ground, ``ready-to-fire'' action that can then be transactionally applied by \daphne.
\begin{Java}
List<Binding> bindings
	= actionElements.getRight().getAvailableBindings();
GroundAction groundAction
	= actionElements.getLeft().bind(bindings.get(0));
\end{Java}
With this ground action at hand, we can check whether it involves service calls in its specification. 
Service calls are responsible for the communication with the outer world and might introduce some fresh data into the system.  In the database, every service call invocation is represented as a textual signature. E.g., the service call \scname{maxAmnt} on employee
 \cname{Kriss} with trip destination \cname{Rome} is internally represented as a string  \cname{maxAmnt(Kriss,Rome)} in the database. 
\daphne knows about this internal representation and offers a special storage for all the service call invocations, i.e., a service call map list. 
This list keeps a map from services to all their invocations in the state. 
Each invocation also comes with the value type that the original service call returns.
\begin{Java}
ServiceCallMapList aSCMList = groundAction.getServiceCallMapList();
\end{Java}
If the evaluation of ground action effects yielded at least one service invocation , one has to instantiate it before updating the database. To do so we create a service manager instance that, given a service call map list, manages corresponding service objects by applying them to processed arguments (extracted from the signatures) and then collecting the results of the  invocations. As soon as the service call map list is fully instantiated, one can finally proceed with executing the ground action
\begin{Java}
ServiceManager sm = new ServiceManager(connection);
if (!aSCMList.equals(null)) {
	aSCMList = sm.getResults(aSCMList,currState.getStateID());}
groundAction.launch(aSCMList);
\end{Java}

\newpage
\section{State-space construction}
\label{sec:ss}
\begin{figure*}[t] 
\centering
\includegraphics[width=.9\textwidth]{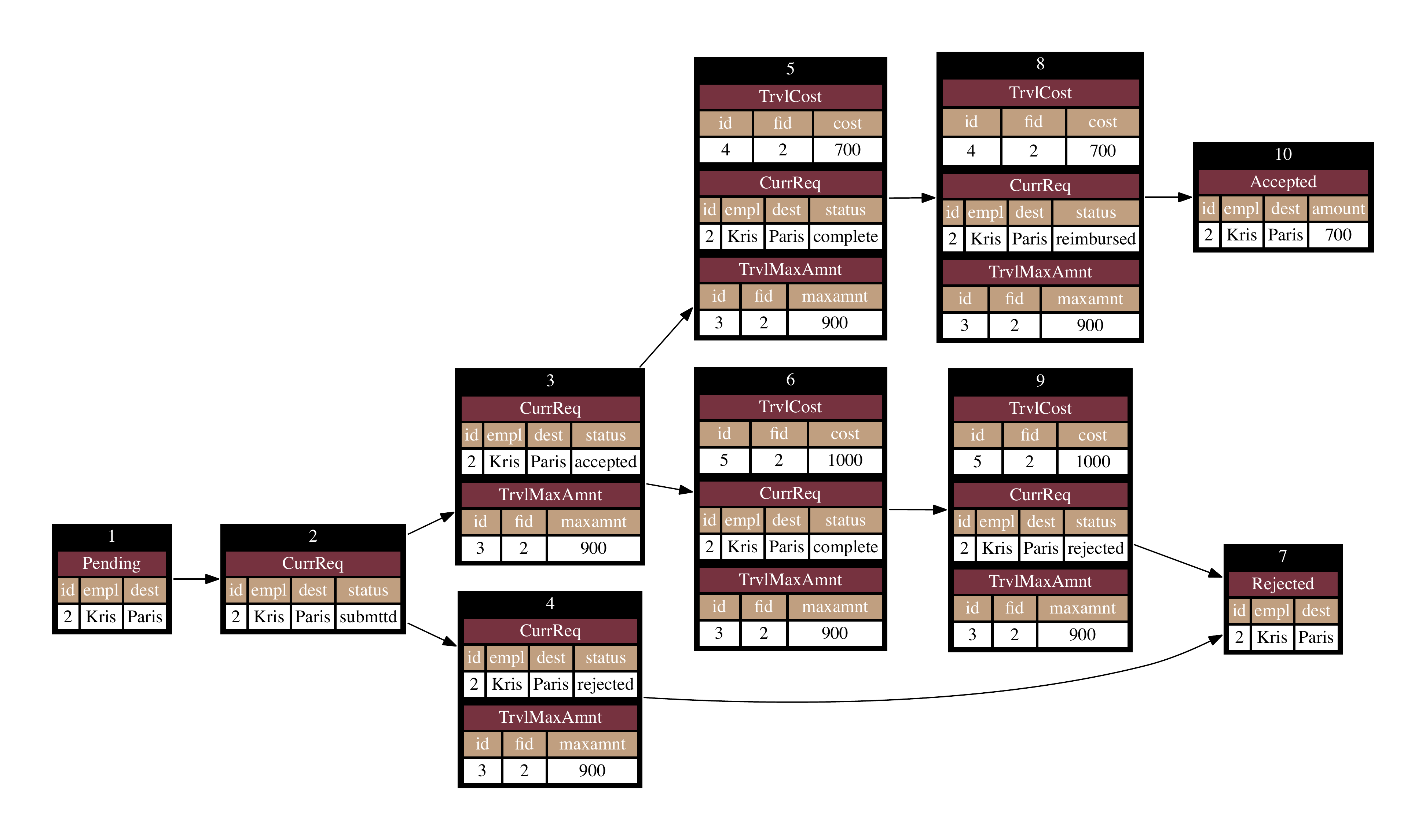}  
\caption{State space of the travel reimbursement process with one pending request}
\label{fig:ss1}
\end{figure*}

A complete example of the state space (cf. Figure~\ref{fig:ss}) of the travel reimbursement with only one pending request $\relname{Pending}(\cname{2}, \cname{Kriss}, \cname{Paris})$ is present in Figure~\ref{fig:ss1}. 
For ease of presentation, we assume that the maximum reimbursable amount \cname{500} and, without loss of generality, we also restrict the range of cost to {400, 600}. Like that one can easily model two scenarios: when a reimbursement request is accepted (i.e., the travel cost is less or equal than the maximum reimbursable amount) and when it is rejected (i.e., the travel cost is greater than the maximum reimbursable amount).

\begin{figure*}[t] 
\centering
\includegraphics[width=.8\textwidth]{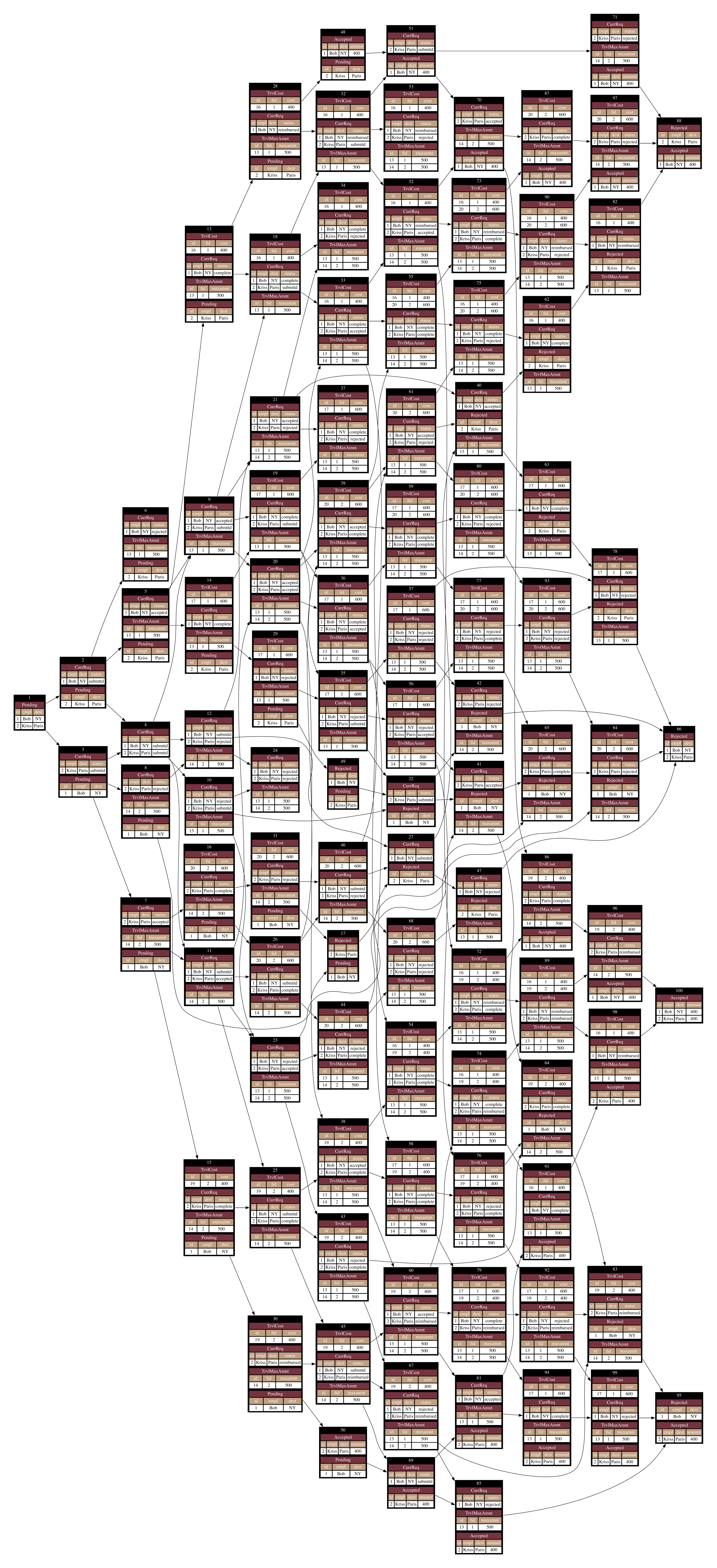}  
\caption{State space of the travel reimbursement process with two pending requests}
\label{fig:ss2}
\end{figure*}

The next example, analogously to Section~\ref{sec:representation} considers two pending requests $\relname{Pending}(\cname{1}, \cname{Bob}, \cname{NY})$ and $\relname{Pending}(\cname{2}, \cname{Kriss}, \cname{Paris})$. The corresponding state space can bee seen in Figure~\ref{fig:ss2}. Given that the travel reimbursement workflow (cf. Figure~\label{fig:ex-bpm}) does not consider any interaction between different process instances, it is enough to see each of such instances evolving ``in isolation'', that is, assuming the same execution scenario (with the same maximum amount an travel cost values) as above. Like that, the state space becomes nothing but a set of all possible combinations of states of two process instances.

%% file: main.bbl
\begin{thebibliography}{10}
\providecommand{\url}[1]{\texttt{#1}}
\providecommand{\urlprefix}{URL }
\providecommand{\doi}[1]{https://doi.org/#1}

\bibitem{BCDDM13}
Bagheri~Hariri, B., Calvanese, D., De~Giacomo, G., Deutsch, A., Montali, M.:
  Verification of relational data-centric dynamic systems with external
  services. In: Proc.\ of PODS (2013)

\bibitem{BCDM14}
Bagheri~Hariri, B., Calvanese, D., Deutsch, A., Montali, M.: State-boundedness
  for decidability of verification in data-aware dynamic systems. In: Proc.\ of
  KR. AAAI Press (2014)

\bibitem{CaDM13}
Calvanese, D., De~Giacomo, G., Montali, M.: Foundations of data aware process
  analysis: {A} database theory perspective. In: Proc.\ of PODS (2013)

\bibitem{CDMP17}
Calvanese, D., {De Giacomo}, G., Montali, M., Patrizi, F.: First-order
  \emph{{\(\mu\)}}-calculus over generic transition systems and applications to
  the situation calculus. Inf.\ and Comp.  \textbf{259}(3),  328--347 (2018).
  \doi{10.1016/j.ic.2017.08.007}

\bibitem{DaHV11}
Damaggio, E., Hull, R., Vacul{\'\i}n, R.: On the equivalence of incremental and
  fixpoint semantics for business artifacts with {Guard-Stage-Milestone}
  lifecycles. In: Proc.\ of BPM (2011)

\bibitem{DDG16}
De~Masellis, R., Di~Francescomarino, C., Ghidini, C., Montali, M., Tessaris,
  S.: Add data into business process verification: Bridging the gap between
  theory and practice. In: Proc.\ of AAAI. AAAI Press (2017)

\bibitem{Dum11}
Dumas, M.: On the convergence of data and process engineering. In: Proc.\ of
  ADBIS. LNCS, vol.~6909. Springer (2011)

\bibitem{FurC85}
Furtado, A.L., Casanova, M.A.: Updating relational views. In: Query Processing
  in Database Systems, pp. 127--142. Springer (1985)

\bibitem{Hull08}
Hull, R.: Artifact-centric business process models: {B}rief survey of research
  results and challenges. In: Proc.\ of OTM. LNCS, vol.~5332. Springer (2008)

\bibitem{KopS16}
K{\"o}pke, J., Su, J.: Towards quality-aware translations of activity-centric
  processes to guard stage milestone. In: Proc.\ of BPM. LNCS, vol.~9850.
  Springer (2016)

\bibitem{KuWR11}
K{\"u}nzle, V., Weber, B., Reichert, M.: Object-aware business processes:
  Fundamental requirements and their support in existing approaches. Int.\ J.\
  of Information System Modeling and Design  \textbf{2}(2) (2011)

\bibitem{LiDV17}
Li, Y., Deutsch, A., Vianu, V.: {VERIFAS:} {A} practical verifier for artifact
  systems. {PVLDB}  \textbf{11}(3) (2017)

\bibitem{MeSW11}
Meyer, A., Smirnov, S., Weske, M.: Data in business processes. Tech. Rep.~50,
  Hasso-Plattner-Institut for IT Systems Engineering, Universit{\"a}t Potsdam
  (2011)

\bibitem{MonC16}
Montali, M., Calvanese, D.: Soundness of data-aware, case-centric processes.
  Int.\ J.\ on Software Tools for Technology Transfer  \textbf{18}(5),
  535--558 (2016)

\bibitem{MonR16}
Montali, M., Rivkin, A.: Model checking {Petri} nets with names using
  data-centric dynamic systems. Formal Aspects of Computing  (2016)

\bibitem{MonR17}
Montali, M., Rivkin, A.: {DB-Nets}: on the marriage of colored {Petri} {Nets}
  and relational databases. Trans.\ on Petri Nets and Other Models of
  Concurrency  \textbf{28}(4) (2017)

\bibitem{Oli07}
Oliv{\'{e}}, A.: Conceptual modeling of information systems. Springer (2007)

\bibitem{Reic12}
Reichert, M.: Process and data: {T}wo sides of the same coin? In: Proc.\ of
  OTM. LNCS, vol.~7565. Springer (2012)

\bibitem{Rich10}
Richardson, C.: Warning: Don't assume your business processes use master data.
  In: Proc.\ of BPM. LNCS, vol.~6336. Springer (2010)

\bibitem{RMRS18}
Ritter, D., Rinderle{-}Ma, S., Montali, M., Rivkin, A., Sinha, A.: Formalizing
  application integration patterns. In: 22nd {IEEE} International Enterprise
  Distributed Object Computing Conference, {EDOC} 2018, Stockholm, Sweden,
  October 16-19, 2018. pp. 11--20 (2018). \doi{10.1109/EDOC.2018.00012}

\bibitem{Reseda18}
Seco, J.C., Debois, S., Hildebrandt, T.T., Slaats, T.: {RESEDA:} declaring live
  event-driven computations as reactive semi-structured data. In: 22nd {IEEE}
  International Enterprise Distributed Object Computing Conference, {EDOC}
  2018, Stockholm, Sweden, October 16-19, 2018. pp. 75--84 (2018).
  \doi{10.1109/EDOC.2018.00020}

\bibitem{SMTD13}
Solomakhin, D., Montali, M., Tessaris, S., De~Masellis, R.: Verification of
  artifact-centric systems: Decidability and modeling issues. In: Proc.\ of
  ICSOC. LNCS, vol.~8274. Springer (2013)

\bibitem{SSWY14}
Sun, Y., Su, J., Wu, B., Yang, J.: Modeling data for business processes. In:
  Proc.\ of ICDE. pp. 1048--1059. {IEEE} Computer Society (2014)

\bibitem{SuSY16}
Sun, Y., Su, J., Yang, J.: Universal artifacts: {A} new approach to business
  process management {(BPM)} systems. ACM TMIS  \textbf{7}(1) (2016)

\bibitem{Vard05}
Vardi, M.Y.: Model checking for database theoreticians. LNCS, vol.~3363, pp.
  1--16. Springer (2005)

\bibitem{XSYY11}
Xu, W., Su, J., Yan, Z., Yang, J., Zhang, L.: An artifact-centric approach to
  dynamic modification of workflow execution. In: Proc.\ of CoopIS. vol.~7044.
  Springer (2011)

\end{thebibliography}
